\documentclass{jpp}
\usepackage{amsmath}
\usepackage{amssymb}	% Math symbols such as \mathbb
\usepackage{bbold}
\usepackage{bm} %allows msans to be used (for matrices' lettering)

\usepackage{xcolor}
\definecolor{myBlue}{RGB}{50,117,180}
\definecolor{myRed}{RGB}{200,40,40}
\definecolor{myGreen}{RGB}{34,139,34}
\definecolor{myLightBlue}{RGB}{193,223,255}
\definecolor{myOrange}{RGB}{255,185,0}

\usepackage{cancel} % cancel terms out of an equation with \cancel{.} or \cancelto{.}{.}
\usepackage[normalem]{ulem} %allows \sout{.} for strikethrough characters
\usepackage{changepage}
\usepackage{float}
\usepackage{mathrsfs}
\usepackage{url}
\usepackage{xfrac}
\usepackage{graphicx}
\usepackage{lipsum} %Lorem ipsum
\graphicspath{{/}}
\usepackage[font=small,labelfont=bf,justification=raggedright,format=plain,singlelinecheck=false]{caption} %justified

\newcommand{\comment}[1]{\ignorespaces} %for inline commenting
\newcommand{\w}[1]{\ensuremath{\mathbf{#1}}} % for vectors (my usual convention is \v but it collides with \hyperref)
\newcommand{\gv}[1]{\ensuremath{\mbox{\boldmath$ #1 $}}} 
\newcommand{\pd}[2]{\frac{\partial #1}{\partial #2}} 

\DeclareMathAlphabet{\mathsfit}{\encodingdefault}{\sfdefault}{m}{n}
\SetMathAlphabet{\mathsfit}{bold}{\encodingdefault}{\sfdefault}{bx}{n}
\newcommand{\abs}[1]{\left| #1 \right|} % for absolute value
\makeatletter
\newcommand*\bigcdot{\mathpalette\bigcdot@{.5}}
\newcommand*\bigcdot@[2]{\mathbin{\vcenter{\hbox{\scalebox{#2}{$\m@th#1\bullet$}}}}}
\makeatother

\makeatletter
\newcommand{\raisemath}[1]{\mathpalette{\raisem@th{#1}}}
\newcommand{\raisem@th}[3]{\raisebox{#1}{$#2#3$}}
\makeatother

\usepackage{accents}

\makeatletter %makes "@" just a letter
\renewcommand*\env@matrix[1][*\c@MaxMatrixCols c]{%
  \hskip -\arraycolsep
  \let\@ifnextchar\new@ifnextchar
  \array{#1}}
\makeatother

\usepackage{tikz}
\usepackage{diagrams}
\usetikzlibrary{cd}

\usepackage{pgfplots}
\usepgfplotslibrary{patchplots}
\usepgfplotslibrary{external}
\usepackage{tikz}
\usetikzlibrary{external}
\usetikzlibrary{calc}
\usetikzlibrary{arrows,shapes,backgrounds,arrows.meta,shapes.arrows,automata,quotes,intersections}
\usetikzlibrary{decorations.text}
\usetikzlibrary{decorations.fractals}
\usetikzlibrary{decorations.pathmorphing}
\usetikzlibrary{decorations.markings}

\usetikzlibrary{decorations.pathreplacing}
\makeatletter
    \let\pgf@decorate@@brace@brace@code@old\pgf@decorate@@brace@brace@code
    \def\pgf@decorate@@brace@brace@code{
        \ifdim\pgfdecoratedremainingdistance<4\pgfdecorationsegmentamplitude
            \pgftransformxscale{\pgfdecoratedremainingdistance/4\pgfdecorationsegmentamplitude}
            \pgfdecoratedremainingdistance=4\pgfdecorationsegmentamplitude
        \fi
        \pgf@decorate@@brace@brace@code@old
    }
\makeatother

\newcommand*\mC{\mathbb{C}}

\newcommand*\mmD{\mathbb{D}}

\newcommand*\mG{\mathbb{G}}

\newcommand*\mM{\mathbb{M}}

\newcommand*\mmP{\mathcal{P}}
\newcommand*\mR{\mathbb{R}}

\newcommand*\mT{\mathcal{T}}

\newcommand*\mX{\mathfrak{X}}

\newcommand*\mg{\mathfrak{g}}

\newcommand*\mZero{\mathbb{0}}

\newcommand*\md{\mathrm{d}}

\DeclareMathAlphabet{\mathpzc}{OT1}{pzc}{m}{it}

\makeatletter
\newcommand*{\defeq}{\mathrel{\rlap{%
                     \raisebox{0.3ex}{$\m@th\cdot$}}%
                     \raisebox{-0.3ex}{$\m@th\cdot$}}%
                     =}
\newcommand*{\eqdef}{=\mathrel{\rlap{%
                     \raisebox{0.3ex}{$\m@th\cdot$}}%
                     \raisebox{-0.3ex}{$\m@th\cdot$}}%
                     }
\makeatother

\DeclareMathOperator*{\argmin}{arg\,min}

\usepackage{mathtools}

\usepackage{hhline}

\usepackage{stmaryrd}

\usepackage{enumitem}

\definecolor{light-gray}{gray}{.85}
\newsavebox{\songboxbox}

\usepackage[most]{tcolorbox}
\newtcbox{\MyBox}[1][red]{on line, size=tight, boxsep=1pt, colframe=#1!50!black, colback=#1!10!white}

\usepackage[framemethod=TikZ]{mdframed}
\mdfsetup{%
    outerlinewidth=1pt,
    innertopmargin=6pt,
    innerbottommargin=6pt,
    skipabove=10pt,
    skipbelow=10pt,
    backgroundcolor=black!10,
    roundcorner=20pt
}

\DeclareFontFamily{U}{matha}{\hyphenchar\font45}
\DeclareFontShape{U}{matha}{m}{n}{
      <5> <6> <7> <8> <9> <10> gen * matha
      <10.95> matha10 <12> <14.4> <17.28> <20.74> <24.88> matha12
      }{}
\makeatletter
\newcommand{\blandor}[1]{\mathbin{\@blandor{#1}}}
\newcommand{\@blandor}[1]{\mathchoice
  {\@@blandor{#1}{\tf@size}}
  {\@@blandor{#1}{\tf@size}}
  {\@@blandor{#1}{\sf@size}}
  {\@@blandor{#1}{\ssf@size}}
}
\newcommand{\@@blandor}[2]{%
    \raisebox{.1ex}{\rotatebox[origin=c]{#1}{%
      \fontsize{#2}{#2}\usefont{U}{matha}{m}{n}\symbol{\string"CE}}}%
}
\makeatother

\usepackage{slashed} %Feynman
\newcommand{\cmmnt}[1]{\ignorespaces} %for inline commenting

\usepackage{environ}
\NewEnviron{eqn}{
\begin{align}\begin{split}
\BODY
\end{split}\end{align}
}

\usepackage{xfrac} 

\usepackage{bbding}
\DeclareFontFamily{U}{MnSymbolC}{}
\DeclareSymbolFont{MnSyC}{U}{MnSymbolC}{m}{n}
\DeclareMathSymbol{\diamondplus}{\mathbin}{MnSyC}{"7C}
\DeclareMathSymbol{\diamonddot}{\mathbin}{MnSyC}{"7E}
\DeclareFontShape{U}{MnSymbolC}{m}{n}{
    <-6>  MnSymbolC5
   <6-7>  MnSymbolC6
   <7-8>  MnSymbolC7
   <8-9>  MnSymbolC8
   <9-10> MnSymbolC9
  <10-12> MnSymbolC10
  <12->   MnSymbolC12}{}

\usepackage{hyperref}

\begin{document}

%\preprint{.}

\title{A gauge-compatible Hamiltonian splitting algorithm for particle-in-cell simulations using finite element exterior calculus} %Title of paper

% actual e-mail address or url should go in the {.}'s for \email and \homepage.


%\author{Alexander S. Glasser}
%\thanks{.}
%\altaffiliation{.}

%\author{Hong Qin}
%\thanks{.}
%\altaffiliation{.}

%\collaboration{.}

%\date{\today}

%%%%%%%%%JPP HEADER%%%%%%%%%%%%%%%%%%%%
\author{Alexander S. Glasser\aff{1,2}
 \corresp{\email{asg5@princeton.edu}},
 \and Hong Qin\aff{1,2}}

\affiliation{
\aff{1}Princeton Plasma Physics Laboratory, Princeton University, Princeton, New Jersey 08543
\aff{2}Department of Astrophysical Sciences, Princeton University, Princeton, New Jersey 08544
}

%\date{\today}

\maketitle %\maketitle must follow title, authors, abstract and \pacs

%%%%%%%%%JPP HEADER (END)%%%%%%%%%%%%%%%%

\hyphenpenalty=1000
\begin{abstract}
A particle-in-cell algorithm is derived with a canonical Poisson structure in the formalism of finite element exterior calculus. The resulting method belongs to the class of gauge-compatible splitting algorithms, which exactly preserve gauge symmetries and their associated conservation laws via the momentum map. We numerically demonstrate this time invariance of the momentum map and its usefulness in establishing precise initial conditions with a desired initial electric field and fixed background charge. The restriction of this canonical, finite element Poisson structure to the 1X2P 1\sfrac{1}{2}-dimensional phase space is also considered and simulated numerically.
\end{abstract}
\hyphenpenalty=50

%\pacs{}% insert suggested PACS numbers in braces on next line

%\maketitle %\maketitle must follow title, authors, abstract and \pacs


%%%%%%%%%%%%%%%%%%%%%%%%%%%%%%%%%%%%%%%%%%%%%%%%%%%
%%%%%%%%%%%%%%%%%%%%%%%%%%%%%%%%%%%%%%%%%%%%%%%%%%%
%%%%%%%%%%%%%%%%%%%%%%%%%%%%%%%%%%%%%%%%%%%%%%%%%%%

\section{Introduction}
Structure-preserving particle-in-cell (PIC) algorithms preserve many of the geometric and topological mathematical structures of a point-particle kinetic plasma model, including its symplectic structure, symmetries, conservation laws, and cohomology \citep{villasenor_rigorous_1992,Esirkepov2001,squire_geometric_2012,evstatiev2013variational,xiao_variational_2013,Moon2015,Qin15JCP,xiao_explicit_2015,he_hamiltonian_2015,crouseilles2015Hamiltonian,qin_canonical_2016,he_hamiltonian_2016,burby2017finite,Morrison2017,kraus_gempic:_2017,kraus_metriplectic_2017,xiao_structure-preserving_2018,xiao_field_2019,glasser_geometric_2020,Hirvijoki2020,Wang2021,xiao_explicit_2021,holderied_mhd-kinetic_2021,Perse2021,oconnor_time_2021,Kormann2021,pinto_variational_2022}. One such structure, gauge symmetry, was first preserved in a PIC algorithm in the Lagrangian formalism via a variational method \citep{squire_geometric_2012}. More recently, it was also demonstrated that gauge symmetry could be exactly preserved in a Hamiltonian PIC algorithm via a gauge-compatible splitting method \citep{glasser_geometric_2020}, or GCSM.

GCSMs are splitting methods whose sub-Hamiltonians ${H_i}$ (satisfying ${H=\sum H_i}$) are each gauge-symmetric and exactly numerically integrable. It can be shown that such methods preserve the momentum map \citep{souriau_structure_1970,marsden_introduction_1999,da_silva_lectures_2001} of a gauge-symmetric Hamiltonian system. In a GCSM PIC algorithm, the momentum map $\mu$ associated with electromagnetic gauge symmetry forms a discrete equivalent of Gauss' law, characterized by the electric field $\w{E}$ and charge density $\rho$ as ${\mu\sim(\nabla\cdot\w{E}-4\pi\rho)/4\pi c}$ (in Gaussian units). Its preservation---${\dot{\mu}=0}$---enforces a discrete local charge conservation law throughout the simulation domain. As we shall see, specifying this momentum map $\mu$ as an initial condition of a plasma simulation furthermore enables the precise assignment of any fixed, background charge that may be desired throughout a simulation.

Most of the literature's recent structure-preserving Hamiltonian PIC methods employ a non-canonical Poisson structure to describe particle position and velocity degrees of freedom ${(\w{X},\w{V})}$ and discrete electromagnetic fields ${(\w{E},\w{B})}$ \citep{Qin15JCP,xiao_explicit_2015,he_hamiltonian_2015,crouseilles2015Hamiltonian,he_hamiltonian_2016,burby2017finite,Morrison2017,kraus_gempic:_2017,kraus_metriplectic_2017,xiao_structure-preserving_2018,xiao_field_2019,xiao_explicit_2021,holderied_mhd-kinetic_2021,Perse2021,Kormann2021,pinto_variational_2022}. This approach hides from view the gauge symmetry of the Vlasov-Maxwell system and the simplicity of its canonical Poisson structure, characterized by the electromagnetic potential $\w{A}$ and its conjugate momentum ${\w{Y}\sim\md\w{A}/\md t}$ as well as particle position and momentum, ${(\w{X},\w{P})}$. The process of `hiding' this gauge symmetry may be formally regarded as the Poisson reduction of the Vlasov-Maxwell system \citep{marsden_reduction_1974,marsden_hamiltonian_1982,marsden_reduction_1986,glasser_geometric_2020}, which strips out gauge symmetry to reduce canonical coordinates ${(\w{A},\w{Y},\w{X},\w{P})}$ to their non-canonical counterparts ${(\w{E},\w{B},\w{X},\w{V})}$.

In this work, we demonstrate the effectiveness of the canonical formalism in simulating the Vlasov-Maxwell system. Using the flexible techniques of finite element exterior calculus \citep{arnold_finite_2006,arnold_finite_2010} previously adopted in \citep{kraus_gempic:_2017}, we discover a canonical finite element Poisson structure for the Vlasov-Maxwell system. From this discrete structure, we derive a gauge-compatible splitting method that defines an exactly solvable symplectic PIC algorithm. We characterize the gauge symmetry of this algorithm and use it to derive the charge-conserving momentum map, $\mu$. We further demonstrate the use of $\mu$ to establish initial conditions with a simple numerical example of a fixed, homogeneous, neutralizing, positive background charge in a Landau damping simulation. Lastly, we consider the restriction of our method to a 1\sfrac{1}{2}-dimensional \citep{crouseilles2015Hamiltonian} 1X2P phase space. We demonstrate that this subsystem retains the essential gauge structure of the Vlasov-Maxwell system, and test its numerical efficacy with a simulation of the Weibel instability \citep{weibel_spontaneously_1959}. 

The remainder of this article is organized as follows: In Section~\ref{momMap}, we briefly introduce the momentum map. In Section~\ref{feecDefine}, we describe aspects of finite element exterior calculus that are used in the algorithm. In Section~\ref{DiscVMStructure}, we derive the canonical Poisson structure of the discrete Vlasov-Maxwell system, its Hamiltonian and its momentum map. In Section~\ref{EOM}, a gauge-compatible splitting method is derived. In Section~\ref{NumResults}, we describe the practical use of the momentum map, and present numerical results from Landau damping and Weibel instability simulations. In Section~\ref{ConcludeSection}, we summarize and conclude.

\section{A Brief Review of the Momentum Map\label{momMap}}

The momentum map \citep{souriau_structure_1970,marsden_introduction_1999,da_silva_lectures_2001} may be viewed as the Hamiltonian manifestation of the Noether principle---i.e., that every smooth symmetry of a dynamical system corresponds to a conserved quantity. We may demonstrate how the momentum map arises from the symmetry transformations of a Lie group on a Poisson manifold, as follows.

Suppose a Poisson manifold $M$ is equipped with symmetry transformations defined by the group action ${\Phi:G\times M\rightarrow M}$, where $G$ is a Lie group whose elements act upon $M$. The group action $\Phi_g$ associated with any fixed ${g\in G}$ is defined such that ${\Phi_g(x)=\Phi(g,x)}$ $\forall$ ${x\in M}$. The corresponding Lie algebra ${\mg=\text{Lie}(G)}$, may be regarded as generating these symmetry transformations, since $\forall$ ${\w{s}\in\mg}$ there exists a vector field ${X_\w{s}\in\mX(M)}$ on $M$ defined by
\begin{eqn}
X_\w{s}=\left.\frac{\md}{\md\epsilon}\right| _{\epsilon=0}\Phi_{\exp(\epsilon\w{s})}.
\label{symmGen}
\end{eqn}
Here, ${\{\exp(\epsilon\w{s})\in G~|~\epsilon\in\mR\}}$ is the one-parameter subgroup of $G$ generated by $\w{s}$. Thus, the vector field $X_\w{s}$ is seen to `infinitesimally generate' the family of symmetry transformations ${\Phi_{\exp(\epsilon\w{s})}}$. Equivalently, one may regard the collection of maps ${\{\Phi_{\exp(\epsilon\w{s})}:M\rightarrow M~|~\epsilon\in\mR\}}$ as the flow of $M$ along the vector field $X_\w{s}$. In this way, each Lie algebra generator ${\w{s}\in\mg}$ corresponds to a generator of transformations on $M$---namely, the vector field $X_\w{s}$.

Now suppose ${\{\cdot,\cdot\}:C^\infty(M)\times C^\infty(M)\rightarrow C^\infty(M)}$ denotes the Poisson bracket on $M$, defining an algebra of smooth functions. The momentum map is then defined as a linear map ${\mu:\mg\rightarrow C^\infty(M)}$ which assigns to every $\w{s}\in\mg$ a generating function ${\mu(\w{s})\in C^\infty(M)}$---henceforth denoted ${\mu_\w{s}=\mu(\w{s})}$---satisfying
\begin{eqn}
\{F,\mu_\w{s}\}=X_\w{s}(F)~~~~\forall~F\in C^\infty(M).
\label{defineMuDerivation}
\end{eqn}
Here, ${X_\w{s}(F)}$ denotes the Lie derivative of $F$ by the vector field $X_\w{s}$. In this way, $\mu_\w{s}$ is a generating function via the Poisson bracket for the symmetry transformation generated by $\w{s}$; in particular, ${X_\w{s}=\{\cdot,\mu_\w{s}\}}$ are equal as derivations on ${C^\infty(M)}$. Therefore, when a momentum map $\mu$ is defined, each ${\w{s}\in\mg}$ corresponds not only to a vector field $X_\w{s}$ on $M$, but to a generating function ${\mu_\w{s}\in C^\infty(M)}$ as well. The linearity of $\mu$ implies that ${\mu_{\w{s}+\w{t}}=\mu_\w{s}+\mu_\w{t}}$, and we more generally denote $\mu$ as the map satisfying ${\mu\cdot\w{s}=\mu_\w{s}}$ $\forall$ ${\w{s}\in\mg}$.

In a symmetric Hamiltonian system, the function $\mu_\w{s}$ is not only a generator of a symmetry transformation, but also a conserved quantity. To see this, consider a symmetric Hamiltonian ${H\in C^\infty(M)}$ that is invariant under a group action $\Phi$, such that
\begin{eqn}
H\circ\Phi_g=H~~~\forall~g\in G.
\label{HamiltonianInvariant}
\end{eqn}
Setting ${g=\exp(\epsilon\w{s})}$ and differentiating both sides of Eq.~(\ref{HamiltonianInvariant}) with respect to $\epsilon$, it follows by Eq.~(\ref{symmGen}) that ${X_\w{s}(H)=0}$ $\forall$ ${\w{s}\in\mg}$. By Eq.~(\ref{defineMuDerivation}), then,
\begin{eqn}
0=\{H,\mu_\w{s}\}=-\dot{\mu}_\w{s}.
\end{eqn}
Therefore, $\mu_\w{s}$ is constant along the flow generated by $H$ on $M$, i.e. it is conserved in time. Since ${0=\dot{\mu}^\w{s}=\md/\md t(\mu\cdot\w{s})=\dot{\mu}\cdot\w{s}}$ $\forall$ ${\w{s}\in\mg}$, we more generally write that ${\dot{\mu}=0}$.

\section{Relevant Aspects of Finite Element Exterior Calculus\label{feecDefine}}

We now introduce notation and briefly describe elementary aspects of finite element exterior calculus \citep{arnold_finite_2006,arnold_finite_2010}, abbreviated FEEC. To approximate differential forms on a smooth manifold $\Omega$ by finite elements, FEEC projects the de Rham complex of differential forms ${0\xrightarrow{\md}\Lambda^0(\Omega)\xrightarrow{\md}\cdots\xrightarrow{\md}\Lambda^n(\Omega)\xrightarrow{\md}0}$ onto spaces of piecewise polynomial differential forms, as depicted in Fig.~\ref{cochainFig}. In particular, given a triangulation of $\Omega$ denoted $\mT_h$ (with maximum diameter $h$ on any given cell), ${\Lambda^p(\mT_h)}$ denotes some finite-dimensional space of piecewise $p$-forms on $\mT_h$ whose coefficients are polynomial over each $p$-cell of $\mT_h$. The projection map ${\pi_h:\Lambda^p(\Omega)\rightarrow\Lambda^p(\mT_h)}$ for each $p$ is required to ensure that the diagram of Fig.~\ref{cochainFig} commutes---in particular, that ${\pi_h\circ\md=\md\circ\pi_h}$. The horizontal arrows of Fig.~\ref{cochainFig} are required to form cochain complexes ($\md\circ\md=0$) such that the projections $\pi_h$ are isomorphisms of cohomology. There exist many possible choices for the finite element spaces ${\Lambda^p(\mT_h)}$, which can vary in their degree of accuracy and with the choice of triangulation $\mT_h$. However, any such choice must ensure that the sequence of spaces ${\cdots\rightarrow\Lambda^p(\mT_h)\rightarrow\cdots}$ constitutes a cochain complex, and that the finite element problem being studied is solvable and well-posed in those spaces.

\begin{figure}
\begin{diagram}
0 & \rTo^{\md} & \Lambda^0(\Omega) & \rTo^{\md} & \cdots & \rTo^{\md} & \Lambda^n(\Omega) & \rTo^{\md} & 0 \\
 & & \dTo^{\pi_h} & & & & \dTo^{\pi_h} & & \\
0 & \rTo^{\md} & \Lambda^0(\mT_h) & \rTo^{\md} & \cdots & \rTo^{\md} & \Lambda^n(\mT_h) & \rTo^{\md} & 0
\end{diagram}
\caption{Given a triangulation $\mT_h$ of the smooth manifold $\Omega$, each space of differential forms ${\Lambda^p(\Omega)}$ of the continuous cochain complex is projected onto a finite element space ${\Lambda^p(\mT_h)}$. There are many possible choices, of varying degrees of accuracy, for the spaces of piecewise polynomial finite elements ${\Lambda^p(\mT_h)}$. The projections $\pi_h$ are required to satisfy ${\pi_h\circ\md=\md\circ\pi_h}$, such that the diagram above is commuting. }
\label{cochainFig}
\end{figure}

For example, a straightforward choice for ${\Lambda^p(\mT_h)}$ is given by the space of all piecewise polynomial $p$-forms of degree ${\leq r}$, denoted ${\mmP_r\Lambda^p(\mT_h)}$. An alternative, more lightweight choice for ${\Lambda^p(\mT_h)}$ is the space of Whitney $p$-forms \citep{whitney_geometric_1957}. These are piecewise linear forms that can be defined using barycentric coordinate functions ${\{\lambda_0,\dots,\lambda_p\}}$ such that for any $p$-simplex $\sigma$, the Whitney $p$-form $\phi_\sigma$ associated with $\sigma$ is given by
\begin{eqn}
\phi_\sigma=\sum\limits_{i=0}^p(-1)^i\lambda_{\sigma(i)}\Big[\md\lambda_{\sigma(0)}\wedge\cdots\wedge\widehat{\md\lambda_{\sigma(i)}}\wedge\cdots\wedge\md\lambda_{\sigma(p)}\Big].
\end{eqn}
(Here the hat signifies that ${\md\lambda_{\sigma(i)}}$ is omitted.) We refer the reader to \citet{arnold_finite_2006,arnold_finite_2010,arnold_finite_2013,arnold_finite_2015} for more detailed descriptions of cochain-complex-conforming finite element spaces.

We may fix notation for calculations with finite elements independent of the choice of space $\Lambda^p(\mT_h)$. For any such choice, we may fix a basis for ${\Lambda^p(\mT_h)}$ with $N_p$ finite elements, and organize them into the ${N_p\times1}$ vector ${\gv{\Lambda}^p}$. The $i^\text{th}$ entry of ${\gv{\Lambda}^p}$ is a basis element we denote ${\Lambda^p_i\in\Lambda^p(\mT_h)}$, which defines a piecewise polynomial $p$-form on $\mT_h$. Any $p$-form ${\w{S}\in\Lambda^p(\mT_h)}$ may thus be expanded in the $\gv{\Lambda}^p$ basis as
\begin{eqn}
\w{S}(\w{x})&=\w{s}\cdot\gv{\Lambda}^p(\w{x})=s_i\Lambda^p_i(\w{x})
\label{basisElement}
\end{eqn}
$\forall$ ${\w{x}\in\abs{\mT_h}}$, (where ${\abs{\mT_h}\subset\mR^n}$ denotes the convex hull of $\mT_h$). The individual components of $\w{S}$ will be denoted
\begin{eqn}
\w{S}(\w{x})_{\mu_1\cdots\mu_p}&=\w{s}\cdot\gv{\Lambda}^p(\w{x})_{\mu_1\cdots\mu_p}=s_i\Lambda^p_i(\w{x})_{\mu_1\cdots\mu_p}.
\label{componentsOfForm}
\end{eqn}
Here, ${\w{s}\in\mR^{N_p}}$ and Einstein summation convention is used for the repeated $i$ index. Greek letters denote coordinate indices. For ${\abs{\mT_h}\subset\mR^3}$, for example, the $\mu^\text{th}$ component of the 1-form basis element $\Lambda^1_i(\w{x})$ is denoted ${\Lambda^1_i(\w{x})_\mu}$ $\forall$ ${\mu\in\{1,2,3\}}$, such that ${\Lambda^1_i(\w{x})=\Lambda^1_i(\w{x})_\mu\md x^\mu}$.

Exterior calculus may be computed in the ${\gv{\Lambda}^0,\dots,\gv{\Lambda}^n}$ bases by simple matrix multiplication. Let us define this explicitly on $\mR^3$, where the exterior derivatives of $\text{0-, 1-,}$ and ${\text{2-forms}}$ can be implemented via matrices that represent the gradient ($\mG$), curl ($\mC$), and divergence ($\mathbb{D}$), respectively. In Table~\ref{dMatrixOps}, we define these matrices to act on the coefficients of forms so as to compute the forms' exterior derivatives. For example, the gradient of the 0-form ${\w{S}=\w{s}\cdot\gv{\Lambda}^0}$ is computed to be ${\md\w{S}=\mG\w{s}\cdot\gv{\Lambda}^1}$, where the matrix ${\mG\in\mR^{N_1\times N_0}}$ is defined such that ${\md\gv{\Lambda}^0=\mG^T\gv{\Lambda}^1}$. This definition gives the desired result since
\begin{eqn}
\md\w{S}=\w{s}\cdot\md\gv{\Lambda}^0=\w{s}\cdot\mG^T\gv{\Lambda}^1=\mG\w{s}\cdot\gv{\Lambda}^1.
\end{eqn}

\begin{table}%[h!]
\centering
\begin{tabular}{c c c c c}
\hline
~~~$p$-Form~~~ & ~~~Abstract $\md$~~~ & ~~~Matrix $\md$~~~ & ~~~Definition~~~ & ~~~Dimensions~~~\\
\hline
$\w{S}=\w{s}\cdot\gv{\Lambda}^0$ & - & - & - & - \\
\hline
$\w{A}=\w{a}\cdot\gv{\Lambda}^1$ & $\w{A}=\md\w{S}$ & $\w{a}=\mG\w{s}$ & $\mG^T\gv{\Lambda}^1=\md\gv{\Lambda}^0$ & ${\mG\in\mR^{N_1\times N_0}}$\\
\hline
$\w{B}=\w{b}\cdot\gv{\Lambda}^2$ & $\w{B}=\md\w{A}$ & $\w{b}=\mC\w{a}$ & $\mC^T\gv{\Lambda}^2=\md\gv{\Lambda}^1$ & ${\mC\in\mR^{N_2\times N_1}}$\\
\hline
$\w{C}=\w{c}\cdot\gv{\Lambda}^3$ & $\w{C}=\md\w{B}$ & $\w{c}=\mmD\w{b}$ & $\mmD^T\gv{\Lambda}^3=\md\gv{\Lambda}^2$ & ${\mmD\in\mR^{N_3\times N_2}}$\\
\hline
\end{tabular}
\caption{The finite element matrix implementation of $\md$ on $\mR^3$. The property ${\md\circ\md=0}$ implies that ${\mC\mG=\mZero}$ and ${\mmD\mC=\mZero}$.}
\label{dMatrixOps}
\end{table}

Lastly, it will be useful to define the \emph{mass matrix} on $\mT_h$ for each basis $\gv{\Lambda}^p$ of finite element $p$-forms. Specifically, we define the mass matrix ${\mM_p\in\mR^{N_p\times N_p}}$ of $\gv{\Lambda}^p$ by
\begin{eqn}
(\mM_p)_{ij}&=\int_{\abs{\mT_h}}\hspace{-3pt}\md\w{x}\Big(\Lambda^p_i,\Lambda^p_j\Big)_p=\int_{\abs{\mT_h}}\Lambda^p_i\wedge\star\Lambda^p_j.
\label{M1def}
\end{eqn}
Here, ${(\cdot,\cdot)_p}$ denotes the inner product on $p$-forms induced by the metric $g_{\mu\nu}$, namely
\begin{eqn}
(\alpha,\beta)_p&=\frac{1}{p!}\alpha_{\mu_1\cdots\mu_p}\beta^{\mu_1\cdots\mu_p}\\
&=\frac{1}{p!}\alpha_{\mu_1\cdots\mu_p}\beta_{\nu_1\cdots\nu_p}g^{\mu_1\nu_1}\cdots g^{\mu_p\nu_p}.
\label{pFormInnerProd}
\end{eqn}
On $\mR^3$, for example, Eq.~(\ref{pFormInnerProd}) defines ${(\alpha,\beta)_p}$ for ${p=1,2}$ simply as the standard inner product ${\alpha\cdot\beta}$. After all, 1- and 2-forms each have three independent components on $\mR^3$ and ${g_{\mu\nu}=\delta_{\mu\nu}}$. We note that ${(\alpha,\beta)_p}$ is symmetric, such that ${\mM_p^T=\mM_p}$. Eq.~(\ref{M1def}) may be understood to define the Hodge star operator $\star$, such that ${\alpha\wedge\star\beta=(\alpha,\beta)_p\md\w{x}}$ for arbitrary $p$-forms $\alpha$ and $\beta$, where $\md\w{x}$ denotes the unique volume form that evaluates to unity on positively oriented vectors that are orthonormal with respect to $g_{\mu\nu}$. The mass matrix will evidently appear wherever metric information is incorporated via the Hodge star.

\section{A Discrete Vlasov-Maxwell Poisson Structure and Hamiltonian\label{DiscVMStructure}}

We now apply the formalism above to define a canonical finite element Poisson structure and Hamiltonian for the Vlasov-Maxwell system. We first describe the Poisson structure of discrete electromagnetic fields on a triangulation $\mT_h$ of the spatial manifold ${\Omega\subset\mR^3}$.

\subsection{The finite element electromagnetic Poisson structure}

To start, let us consider electromagnetic fields on the continuous manifold $\Omega$, using the temporal gauge wherein the electric potential vanishes, ${\phi=0}$. The configuration space for such fields is the set ${Q=\{\w{A}~|~\w{A}\in\Lambda^1(\Omega)\}}$ of possible vector potentials, defined as differential ${\text{1-forms}}$ over $\Omega$. To find a variable conjugate to $\w{A}(\w{x})$, we compute the following variational derivative of the electromagnetic Lagrangian expressed in Gaussian units,
\begin{eqn}
\hspace{-2pt}L_\text{EM}=\frac{1}{8\pi}\int\hspace{-3pt}\md{\w{x}}\hspace{-2pt}\left(\abs{-\frac{1}{c}\dot{\w{A}}}^2-\abs{\md\w{A}}^2\right),~\w{Y}=\frac{\delta L_\text{EM}}{\delta\dot{\w{A}}}=\frac{\dot{\w{A}}}{4\pi c^2}.
\label{maxwellLagrangian}
\end{eqn}
Clearly, ${\w{Y}\in\Lambda^1(\Omega)}$ is also a ${\text{1-form}}$ over $\Omega$ corresponding to negative the electric field, ${\w{Y}=-\w{E}/4\pi c}$. As in Eq.~(\ref{pFormInnerProd}), ${\abs{\alpha}^2=(\alpha,\alpha)_p}$ in Eq.~(\ref{maxwellLagrangian}) denotes the standard inner product on $\mR^3$ for ${p=1,2}$.

The full phase space is then given by the cotangent bundle ${T^*Q=\{\w{A},\w{Y}\}}$ whose canonical symplectic structure is defined by the Poisson bracket \citep{marsden_hamiltonian_1982}
\begin{eqn}
\{F,G\}=\int\left(\frac{\delta F}{\delta\w{A}}\cdot\frac{\delta G}{\delta\w{Y}}-\frac{\delta G}{\delta\w{A}}\cdot\frac{\delta F}{\delta\w{Y}}\right)\md\w{x}.
\label{continuousPoissonBracket}
\end{eqn}
Here, ${F[\w{A},\w{Y}]}$ and ${G[\w{A},\w{Y}]}$ are arbitrary functionals on ${T^*Q}$.

We now map this geometric description of fields on $\Omega$ to its triangulation $\mT_h$. On $\mT_h$, the fields $\w{A}$ and $\w{Y}$ can be defined by their expansion in the corresponding basis for finite element ${\text{1-forms}}$:
\begin{eqn}
\w{A}(t,\w{x})&=\w{a}(t)\cdot\gv{\Lambda}^1(\w{x})\\
\w{Y}(t,\w{x})&=\w{y}(t)\cdot\mM_1^{-1}\cdot\gv{\Lambda}^1(\w{x}).
\label{AandYDefine}
\end{eqn}
Here, ${\gv{\Lambda}^1}$ is an ${N_1\times1}$ vector of basis elements and ${\w{a},\w{y}\in\mR^{N_1}}$ denote coefficients, as in Eq.~(\ref{basisElement}). $\w{a}$ and $\w{y}$ are identified as dynamical variables by explicitly notating their time dependence. The inverse factor of the 1-form mass matrix $\mM_1$ in Eq.~(\ref{AandYDefine}) follows from computing the conjugate momentum of $\w{a}$ in the following discretization of $L_\text{EM}$:
\begin{eqn}
\hspace{-2pt}L_\text{EM}&=\frac{1}{8\pi}\int_{\abs{\mT_h}}\hspace{-3pt}\md{\w{x}}\hspace{-2pt}\left(\abs{-\frac{1}{c}\dot{\w{a}}\cdot\gv{\Lambda}^1}^2-\abs{\w{a}\cdot\md\gv{\Lambda}^1}^2\right)\\
&=\frac{1}{8\pi}\int_{\abs{\mT_h}}\hspace{-3pt}\md{\w{x}}\hspace{-2pt}\left(\frac{1}{c^2}\dot{\w{a}}\cdot\mM_1\cdot\dot{\w{a}}-\w{a}\cdot\mC^T\mM_2\mC\cdot\w{a}\right),~\w{y}=\frac{\delta L_\text{EM}}{\delta\dot{\w{a}}}=\frac{\mM_1\dot{\w{a}}}{4\pi c^2}
\label{discMaxwellLagrangian}
\end{eqn}
where we have substituted ${\md\gv{\Lambda}^1=\mC^T\gv{\Lambda}^2}$ from ${\text{Table }\ref{dMatrixOps}}$ and used mass matrices as defined in Eq.~(\ref{M1def}). Comparing Eqs.~(\ref{maxwellLagrangian}) and Eq.~(\ref{discMaxwellLagrangian}) yields the desired expansion for $\w{Y}$ in Eq.~(\ref{AandYDefine}). (This inverse factor of $\mM_1$ will further ensure that the Poisson bracket of Eq.~(\ref{NewBracket2}) is canonical.)

To discretize the Poisson bracket of Eq.~(\ref{continuousPoissonBracket}), we must first express ${\delta F/\delta\w{A}}$ in terms of ${\partial F/\partial\w{a}}$ for a discrete variation ${\delta\w{A}=\delta\w{a}\cdot\gv{\Lambda}^1}$.  Since variational derivatives are valued dually to their variations, following \citet{kraus_gempic:_2017} it is appropriate to require that 
\begin{eqn}
{\left\langle\frac{\delta F}{\delta\w{A}},\delta\w{a}\cdot\gv{\Lambda}^1\right\rangle}_{L^2\Lambda^1}={\left\langle\frac{\partial F}{\partial\w{a}},\delta\w{a}\right\rangle}_{\mR^{N_1}}.
\label{equalInnerProducts}
\end{eqn}
Here, ${\langle\cdot,\cdot\rangle_{L^2\Lambda^1}=\int_{\abs{\mT_h}}\md\w{x}(\cdot,\cdot)_{p=1}}$ denotes the ${L^2\Lambda^1}$ inner product, and ${\langle\cdot,\cdot\rangle_{\mR^{N_1}}}$ the standard inner product on $\mR^{N_1}$. In particular, setting $\delta a_j=\delta_{ij}$ for some fixed, arbitrary ${i\in[1,N_1]}$ and using ${(\cdot,\cdot)_p}$ as defined in Eq.~(\ref{pFormInnerProd}), Eq.~(\ref{equalInnerProducts}) yields
\begin{eqn}
\int_{\abs{\mT_h}}\hspace{-3pt}\md\w{x}\left(\frac{\delta F}{\delta\w{A}},\Lambda^1_i(\w{x})\right)_{p=1}=\frac{\partial F}{\partial a_i}.
\label{firstStepVariation}
\end{eqn}
To solve for ${\delta F/\delta\w{A}}$, we expand ${\delta F/\delta\w{A}=\w{f}\cdot\gv{\Lambda}^1}$ in the ${\gv{\Lambda}^1}$ basis for some ${\w{f}\in\mR^{N_1}}$, such that evaluating Eq.~(\ref{firstStepVariation}) implies ${\w{f}=\mM_1^{-1}\cdot\partial F/\partial\w{a}}$. Therefore,
\begin{eqn}
\frac{\delta F}{\delta\w{A}}&=\frac{\partial F}{\partial\w{a}}\cdot\mM_1^{-1}\cdot\gv{\Lambda}^1.
\label{dFA_Equation}
\end{eqn}
The discrete variation ${\delta\w{Y}=\delta\w{y}\cdot\mM_1^{-1}\cdot\gv{\Lambda}^1}$ establishes a similar result for $\w{Y}$, namely
\begin{eqn}
\frac{\delta F}{\delta\w{Y}}=\frac{\partial F}{\partial\w{y}}\cdot\gv{\Lambda}^1.
\label{dFY_Equation}
\end{eqn}

Thus, to derive the discrete Poisson bracket, we substitute Eqs.~(\ref{dFA_Equation}-\ref{dFY_Equation}) into Eq.~(\ref{continuousPoissonBracket}) and integrate over $\abs{\mT_h}$ to find:
\begin{eqn}
\{F,G\}=\pd{F}{\w{a}}\cdot\pd{G}{\w{y}}-\pd{G}{\w{a}}\cdot\pd{F}{\w{y}}.
\label{NewBracket2}
\end{eqn}
True to its continuous counterpart in Eq.~(\ref{continuousPoissonBracket}), the discrete Poisson bracket of Eq.~(\ref{NewBracket2}) is in canonical symplectic form (i.e. Darboux coordinates).

\subsection{The finite element Vlasov-Maxwell system}

The full Poisson structure of the discrete Vlasov-Maxwell system readily follows from the foregoing analysis of its electromagnetic subsystem. To describe a system of $L$ particles, we let ${\w{X}_\ell,\w{P}_\ell\in\mR^3}$ ${\ell\in[1,L]}$ denote the position and momentum of the $\ell^\text{th}$ particle and let ${m_\ell}$ and ${q_\ell}$ denote its mass and charge. Particles may then be characterized by a Klimontovich distribution, ${f(\w{x},\w{p})=\sum_\ell\delta(\w{x}-\w{X}_\ell)\delta(\w{p}-\w{P}_\ell)}$. Particle phase space is defined as usual with a canonical bracket on position ${\w{X}_\ell}$ and momentum ${\w{P}_\ell}$.

Combining Eq.~(\ref{NewBracket2}) with the Poisson bracket for these $L$ particles, therefore, the discrete Vlasov-Maxwell Poisson structure is given by
\begin{eqn}
\{F,G\}&=\pd{F}{\w{a}}\cdot\pd{G}{\w{y}}-\pd{G}{\w{a}}\cdot\pd{F}{\w{y}}+\sum\limits_{\ell=1}^L\left(\frac{\partial F}{\partial\w{X}_\ell}\cdot\frac{\partial G}{\partial\w{P}_\ell}-\frac{\partial G}{\partial\w{X}_\ell}\cdot\frac{\partial F}{\partial\w{P}_\ell}\right).
\label{discretePoissonBracket}
\end{eqn}
Here, $F$ and $G$ are arbitrary functionals on the discrete Poisson manifold ${(M_d,\{\cdot,\cdot\})}$, where each point ${m_d\in M_d}$ is defined by the data
\begin{eqn}
m_d&=(\w{a},\w{y},\w{X}_1,\dots,\w{X}_L,\w{P}_1,\dots,\w{P}_L)\\&\in\mR^{N_1}\times\mR^{N_1}\times\mR^{3L}\times\mR^{3L}.
\label{phaseSpacePt}
\end{eqn}
We now define a Hamiltonian ${H_\text{VM}=H_\text{VM}[\w{a},\w{y},\w{X}_i,\w{P}_i]}$ on $M_d$, given in Gaussian units by
\begin{eqn}
H_\text{VM}=H_\text{EM}+&H_\text{Kinetic}\\
\text{where}~~~H_\text{EM}&=\frac{1}{8\pi}\int_{\abs{\mT_h}}\hspace{-3pt}\md\w{x}\left(\abs{-4\pi c\w{Y}}^2+\abs{\md\w{A}}^2\right)\\
&=\frac{1}{8\pi}\bigg((4\pi c)^2\w{y}\cdot\mM_1^{-1}\cdot\w{y}+\w{a}\cdot\mC^T\mM_2\mC\cdot\w{a}\bigg)\\
H_\text{Kinetic}&=\sum\limits_{\ell=1}^L\frac{1}{2m_\ell}\Big|\w{P}_\ell-\frac{q_\ell}{c}\w{A}(\w{X}_\ell)\Big|^2
\label{discHamiltonian}
\end{eqn}
where ${\w{A}(\w{X}_\ell)=\w{a}\cdot\gv{\Lambda}^1(\w{X}_\ell)}$. In $H_\text{EM}$, we have substituted Eqs.~(\ref{M1def}) and (\ref{AandYDefine}). The difference in $H_\text{Kinetic}$ is taken componentwise, i.e. ${P_{\ell\mu}-(q_\ell/c)\w{a}\cdot\gv{\Lambda}^1(\w{X}_\ell)_\mu}$ for ${\mu\in\{1,2,3\}}$. Hereafter, we denote components of $\w{P}_\ell$ by $P_{\ell\mu}$ and those of $\w{X}_\ell$ by $X_\ell^\mu$.

\subsection{Gauge structure\label{GaugeStruct}}

We now examine the gauge structure of this discrete Vlasov-Maxwell system and derive its corresponding momentum map. We first note that because ${\mC\mG=0}$, $H_\text{VM}$ of Eq.~(\ref{discHamiltonian}) is invariant under any gauge transformation ${\Phi_{\exp(\w{s})}:M_d\rightarrow M_d}$ of the form
\begin{eqn}
\Phi_{\exp(\w{s})}\left(\begin{matrix}
\w{a}\\
\w{y}\\
\w{X}_\ell\\
\w{P}_\ell
\end{matrix}\right)=\left(\begin{matrix}
\w{a}+\mG\w{s}\\
\w{y}\\
\w{X}_\ell\\
\w{P}_\ell+\frac{q_\ell}{c}\mG\w{s}\cdot\gv{\Lambda}^1(\w{X}_\ell)\end{matrix}\right)
\label{gaugeTransf}
\end{eqn}
$\forall~{\ell\in[1,L]}$, ${\w{s}\in\mR^{N_0}}$. Since ${\Phi_{\exp(\w{s})}\circ\Phi_{\exp(\w{t})}=\Phi_{\exp(\w{s}+\w{t})}}$, such transformations form an abelian group. Evidently, they are also generated by vector fields ${X_\w{s}\in\mX(M_d)}$ of the form
\begin{eqn}
X_\w{s}=\left.\frac{\md}{\md\epsilon}\right|_{\epsilon=0}\Phi_{\exp(\epsilon\w{s})}=\mG_{jk}s_k\left[\partial_{a_j}+\sum\limits_{\ell=1}^L\frac{q_\ell}{c}\Lambda^{1}_j(\w{X}_\ell)_\mu\partial_{P_{\ell\mu}}\right]
\label{vsGenerator}
\end{eqn}
expressed using Einstein summation convention.

We may check whether $\Phi$ of Eq.~(\ref{gaugeTransf}) is a canonical transformation, that is, whether it preserves the Poisson bracket of Eq.~(\ref{discretePoissonBracket})---
\begin{eqn}
\{F,G\}\circ\Phi_{\exp(\w{s})}=\{F\circ\Phi_{\exp(\w{s})},G\circ\Phi_{\exp(\w{s})}\}
\end{eqn}
---$\forall$ $\w{s}$. It suffices to check this condition infinitesimally, i.e. whether
\begin{eqn}
X_\w{s}(\{F,G\})&=\{X_\w{s}(F),G\}+\{F,X_\w{s}(G)\}.
\label{infinitesimalGroupAction}
\end{eqn}
After canceling terms by the equality of mixed partials, verifying Eq.~(\ref{infinitesimalGroupAction}) reduces to checking that
\begin{eqn}
\hspace{-15pt}0&=\sum\limits_{\ell=1}^L\frac{q_\ell}{c}\mG_{jk}s_k\Big[\partial_{X_\ell^\nu}\Lambda^{1}_j(\w{X}_\ell)_\mu-\partial_{X_\ell^\mu}\Lambda^{1}_j(\w{X}_\ell)_\nu\Big](\partial_{P_{\ell\mu}}F)(\partial_{P_{\ell\nu}}G)\\
&=\sum\limits_{\ell=1}^L\frac{q_\ell}{c}\Big[\mG\w{s}\cdot\md\gv{\Lambda}^1(\w{X}_\ell)_{\nu\mu}\Big](\partial_{P_{\ell\mu}}F)(\partial_{P_{\ell\nu}}G).
\label{remainderTerm}
\end{eqn}
Each term in this sum is seen to vanish, however, since
\begin{eqn}
\mG\w{s}\cdot\md\gv{\Lambda}^1=\w{s}\cdot\mG^T\md\gv{\Lambda}^1=\w{s}\cdot\md(\mG^T\gv{\Lambda}^1)=\w{s}\cdot\md\md\gv{\Lambda}^0=0.
\end{eqn}
Thus, $X_\w{s}$ generates a canonical group action, as desired.

We now find the momentum map $\mu$ of this canonical group action by solving for its generating functions. In particular, given any ${\w{s}\in\mR^{N_0}}$, we seek a generating function $\mu_\w{s}$ such that ${X_\w{s}=\{\cdot,\mu_\w{s}\}}$ as derivations, as in Eq.~(\ref{defineMuDerivation}). By comparing Eqs.~(\ref{discretePoissonBracket}) and (\ref{vsGenerator}), we see that ${X_\w{s}=\{\cdot,\mu_\w{s}\}}$ holds if and only if
\begin{eqn}
\frac{\partial\mu_\w{s}}{\partial\w{a}}&=0\\
\frac{\partial\mu_\w{s}}{\partial\w{y}}&=\mG\w{s}\\
\frac{\partial\mu_\w{s}}{\partial\w{X}_\ell}&=-\frac{q_\ell}{c}\mG\w{s}\cdot\gv{\Lambda}^1(\w{X}_\ell)\\
\frac{\partial\mu_\w{s}}{\partial\w{P}_\ell}&=0~~~\forall~{\ell\in[1,L]}.
\label{pdeForMuS}
\end{eqn}
Since ${\mG^T\gv{\Lambda}^1=\md\gv{\Lambda}^0}$ is an exact form, this linear system of partial differential equations is readily solved by
\begin{eqn}
\mu_\w{s}=\mG\w{s}\cdot\w{y}-\sum\limits_{\ell=1}^L\frac{q_\ell}{c}\w{s}\cdot\gv{\Lambda}^0(\w{X}_\ell).
\end{eqn}

The momentum map $\mu$ characterizing all generating functions $\{\mu_\w{s}\}$ is defined by requiring that ${\mu\cdot\w{s}=\mu_\w{s}}$ $\forall$ $\w{s}$. Therefore, $\mu$ is given by
\begin{eqn}
\mu=\mG^T\w{y}-\sum\limits_{\ell=1}^L\frac{q_\ell}{c}\gv{\Lambda}^0(\w{X}_\ell).
\label{MomMapGaussLaw}
\end{eqn}
Setting ${\mu=0}$, Eq.~(\ref{MomMapGaussLaw}) is a discrete form of Gauss' law,
\begin{eqn}
0=(\nabla\cdot\w{E}-4\pi\rho)/4\pi c
\end{eqn}
where ${\w{E}=-4\pi c\w{Y}}$ and ${\rho=\sum_\ell q_\ell\gv{\Lambda}^0(\w{X}_\ell)}$.

Any nonzero value of $\mu$ indicates that the divergence of the electric field is not entirely accounted for by the dynamical particles labeled ${\ell\in[1,L]}$. Since ${\dot{\mu}=0}$, such a nonzero $\mu$ acts as a fixed, external, nondynamical background charge that persists throughout a simulation, and in a manner that remains entirely structure-preserving. In particular, we may regard $\mu$ as representing an external charge density,
\begin{eqn}
\mu\sim\rho_\text{ext}/c.
\label{interpretMomMap}
\end{eqn}
As we shall demonstrate, Eq.~(\ref{interpretMomMap}) can be useful in establishing precise initial conditions in a PIC simulation.

\section{Equations of Motion\label{EOM}}

\subsection{Continuous-time equations of motion}

Let us now derive equations of motion via the Hamiltonian of Eq.~(\ref{discHamiltonian}) and the Poisson bracket of Eq.~(\ref{discretePoissonBracket}). We find:
\begin{eqn}
\dot{X}_\ell^\mu&=\{X_\ell^\mu,H_\text{VM}\}=\frac{1}{m_\ell}\Big(P_{\ell\mu}-\frac{q_\ell}{c}\w{A}(\w{X}_\ell)_\mu\Big)\\
\dot{P}_{\ell\mu}&=\{P_{\ell\mu},H_\text{VM}\}=\frac{q_\ell}{c}\dot{X}_\ell^\nu\partial_{X_\ell^\mu}\w{A}(\w{X}_\ell)_\nu\\
\dot{\w{a}}&=\{\w{a},H_\text{VM}\}=4\pi c^2\mM_1^{-1}\cdot\w{y}\\
\dot{\w{y}}&=\{\w{y},H_\text{VM}\}=-\frac{1}{4\pi}\mC^T\mM_2\mC\cdot\w{a}+\sum\limits_{\ell=1}^L\frac{q_\ell}{c}\dot{X}_\ell^\mu\gv{\Lambda}^1(\w{X}_\ell)_\mu
\label{theEOM}
\end{eqn}
where ${\w{A}(\w{X}_\ell)_\mu=\w{a}\cdot\gv{\Lambda}^1(\w{X}_\ell)_\mu}$ using the component notation of Eq.~(\ref{componentsOfForm}).

We remark that these equations of motion allow us to reexpress the conservation of the momentum map, ${\dot{\mu}=0}$, as a discrete, local charge conservation law in conservative form. In particular, we may take the time derivative $\dot{\mu}$ of Eq.~(\ref{MomMapGaussLaw}) and substitute $\dot{\w{y}}$ of Eq.~(\ref{theEOM}), noting that ${\mC\mG=0}$ to find
\begin{eqn}
(\mG^T\w{j}-\dot{\rho})/c=0
\label{consFormChargeConsLaw}
\end{eqn}
where ${\rho=\sum_\ell q_\ell\gv{\Lambda}^0(\w{X}_\ell)}$ and ${\w{j}=\sum_\ell q_\ell\dot{X}^\mu_\ell\gv{\Lambda}^1(\w{X}_\ell)_\mu}$. As in Eq.~(\ref{MomMapGaussLaw}), we note a correspondence between the discrete and continuous operators ${\mG^T\sim(-\nabla\cdot)}$. Thus, Eq.~(\ref{consFormChargeConsLaw}) is a discrete equivalent of the charge conservation law ${\partial_t{\rho}+\nabla\cdot\w{j}=0}$, as seen in \citet{kraus_gempic:_2017}.

Eq.~(\ref{theEOM}) is sometimes referred to as a semi-discrete system, since it describes a discretely defined system evolving in continuous time. We now proceed to a fully discrete, algorithmic system by defining a gauge-compatible splitting method.

\subsection{Discrete-time equations of motion via a gauge-compatible splitting}
Using the Vlasov-Maxwell splitting discovered in \citet{he_hamiltonian_2015} and adapted to canonical coordinates in \citet{glasser_geometric_2020}, we split $H_\text{VM}$ of Eq.~(\ref{discHamiltonian}) into five sub-Hamiltonians, as follows:
\begin{eqn}
H_\text{VM}=H_\w{A}+H_\w{Y}&+H^x_\text{Kinetic}+H^y_\text{Kinetic}+H^z_\text{Kinetic}\\
\text{where}~~~~~~~~H_\w{A}&=\frac{1}{8\pi}\w{a}\cdot\mC^T\mM_2\mC\cdot\w{a}\\
H_\w{Y}&=\frac{1}{8\pi}(4\pi c)^2\w{y}\cdot\mM_1^{-1}\cdot\w{y}\\
H^\alpha_\text{Kinetic}&=\sum\limits_{\ell=1}^L\frac{1}{2m_\ell}\Big(P_{\ell\alpha}-\frac{q_\ell}{c}\w{A}(\w{X}_\ell)_\alpha\Big)^2
\label{splitHamiltonian}
\end{eqn}
$\forall$ ${\alpha\in\{x,y,z\}}$. We immediately observe that each sub-Hamiltonian remains invariant under the gauge transformation $\Phi_{\exp(\w{s})}$ defined in Eq.~(\ref{gaugeTransf}).

Let us now examine the equations of motion of each subsystem, omitting equations for the subsystems' static degrees of freedom:
\begin{eqn}
H_\w{A}:&\hspace{11.5pt}\dot{\w{y}}\hspace{15.5pt}=-\frac{1}{4\pi}\mC^T\mM_2\mC\cdot\w{a}\\
H_\w{Y}:&\hspace{11.5pt}\dot{\w{a}}\hspace{15.5pt}=4\pi c^2\mM_1^{-1}\cdot\w{y}\vspace{4pt}\\
\hspace{30pt}H_{\text{Kinetic}}^\alpha:&
\begin{cases}
\dot{X}_\ell^\alpha&=\frac{1}{m_\ell}\Big(P_{\ell\alpha}-\frac{q_\ell}{c}\w{A}(\w{X}_\ell)_\alpha\Big)\vspace{4pt}\\
\dot{P}_{\ell\mu}&=\frac{q_\ell}{c}\dot{X}_\ell^\alpha\partial_{X_\ell^\mu}\w{A}(\w{X}_\ell)_\alpha\vspace{4pt}\\
\dot{\w{y}}&=\sum\limits_{\ell=1}^L\frac{q_\ell}{c}\dot{X}_\ell^\alpha\gv{\Lambda}^1(\w{X}_\ell)_\alpha.
\end{cases}
\label{SubsystemEOM}
\end{eqn}
To clarify notation in Eq.~(\ref{SubsystemEOM}), we emphasize that, in the $H_{\text{Kinetic}}^\alpha$ subsystem, ${\dot{X}_\ell^\mu=0}$ for ${\mu\neq\alpha}$. (Here, $\alpha$ is regarded as fixed while $\mu$ ranges over all $\{{x,y,z}\}$ indices.) Thus, the equations of motion ${X}_\ell^{\mu\neq\alpha}$ are omitted above.

Furthermore, it follows from a simple calculation that ${\ddot{X}_\ell^\alpha=0}$ in $H_{\text{Kinetic}}^\alpha$ so that ${\dot{X}_\ell^\alpha}$ is constant during the evolution of each subsystem. As a result, all subsystems above are exactly integrable. Eq.~(\ref{SubsystemEOM}) therefore defines a gauge-compatible splitting method.

More concretely, an evolution over the timestep $[t,t+\Delta]$ in each subsystem is fully specified by
\begin{eqn}
H_\w{A}:&\hspace{15pt}\w{y}(t+\Delta)=\w{y}(t)-\frac{\Delta}{4\pi}\mC^T\mM_2\mC\cdot\w{a}(t)\vspace{4pt}\\
H_\w{Y}:&\hspace{15pt}\w{a}(t+\Delta)=\w{a}(t)+\Delta 4\pi c^2\mM_1^{-1}\cdot\w{y}(t)\\
H_{\text{Kinetic}}^\alpha:&\\
&\hspace{-55pt}
\begin{cases}
X_\ell^\alpha(t+\delta)&\hspace{-7pt}=X_\ell^\alpha(t)+\frac{\delta}{m_\ell}\Big(P_{\ell\alpha}(t)-\frac{q_\ell}{c}\w{a}(t)\cdot\gv{\Lambda}^1(\w{X}_\ell(t))_\alpha\Big)\\
P_{\ell\mu}(t+\Delta)&\hspace{-7pt}=P_{\ell\mu}(t)+\frac{q_\ell}{c}\dot{X}_\ell^\alpha(t)\w{a}(t)\cdot\smashoperator{\int\limits_t^{t+\Delta}}\hspace{-4pt}\md t'\partial_{X_\ell^\mu(t')}\gv{\Lambda}^1(\w{X}_\ell(t'))_\alpha\\
\w{y}(t+\Delta)&\hspace{-7pt}=\w{y}(t)+\sum\limits_{\ell=1}^L\frac{q_\ell}{c}\dot{X}_\ell^\alpha(t)\smashoperator{\int\limits_t^{t+\Delta}}\hspace{-4pt}\md t'\gv{\Lambda}^1(\w{X}_\ell(t'))_\alpha.
\end{cases}
\label{integratedEvolution}
\end{eqn}
In Eq.~(\ref{integratedEvolution}), $t$ is a fixed initial time and ${\delta\in[0,\Delta]}$ parametrizes the particle trajectory ${\w{X}_\ell(t)\rightarrow\w{X}_\ell(t+\Delta)}$ during one timestep of the ${H_{\text{Kinetic}}^\alpha}$ subsystem, which forms a straight line segment in the $\hat{\alpha}$ direction. Since $\gv{\Lambda}^1$ is comprised of piecewise polynomial differential $\text{1-forms}$, $\gv{\Lambda}^1$ and its derivatives are integrable in closed form along the straight path ${\w{X}_\ell(t+\delta)}$. Thus, Eq.~(\ref{integratedEvolution}) defines a symplectic algorithm---specifically a gauge-compatible splitting method---that can be computed exactly.

\section{Numerical Examples\label{NumResults}}

We now demonstrate the efficacy of this algorithm numerically. In Section~\ref{LandauDamping}, we first consider a one-dimensional simulation of Landau damping electrons, choosing simulation parameters similar to those of \citet{xiao_explicit_2015}, against a fixed, homogeneous, positive background charge. Then, in Section~\ref{1X2PSect}, we reexpress the 1X2V simulation approach pursued in \citet{crouseilles2015Hamiltonian} and \citet{kraus_gempic:_2017} in canonical coordinates, thereby deriving the restriction of our algorithm to 1X2P phase space---comprised of one spatial dimension and two dimensions of canonical momentum. Finally, in Section~\ref{WeibelInstability}, we simulate the electromagnetic Weibel instability in this restricted phase space.

\subsection{Landau Damping\label{LandauDamping}}

Using Whitney form finite elements, a 650-cell domain $\mT_h$ with periodic boundaries is constructed. Each cell is assigned width ${w_x=2.4\times10^{-2}\text{ cm}}$, and ${26\times10^6}$ electrons are simulated (40,000 per cell, when unperturbed). With electron temperature at ${T_e=5}$ keV, the setup has Debye length ${\lambda_D=1.0\text{ cm}}$ and plasma frequency ${\omega_p=3.0\times10^9\text{ rad/s}}$, roughly mirroring physical parameters of \citet{xiao_explicit_2015}.

\begin{figure}
\begin{center}
\includegraphics[width=.7\linewidth]{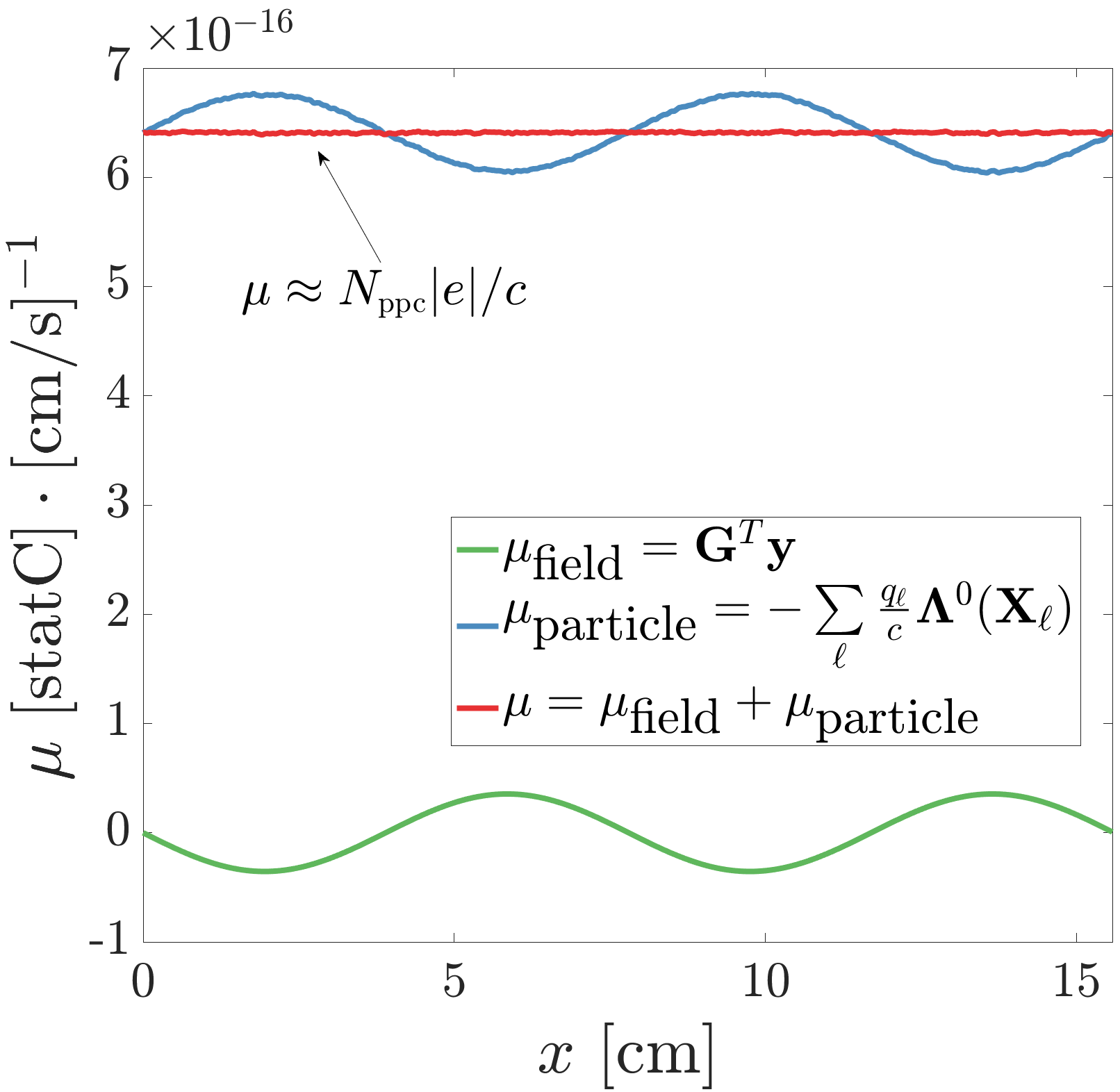}
\end{center}
\caption{The terms of Eq.~(\ref{MomMapGaussLaw}) are plotted over the simulation domain at time ${t=0}$, characterizing initial conditions by the momentum map ${\mu=\mu_\text{field}+\mu_\text{particle}}$.}
\label{MomentumMaps}
\end{figure}

We assume the electric field to have initial perturbation at ${t=0}$
\begin{eqn}
\w{Y}=-\frac{\w{E}}{4\pi c}=-\frac{E_0\cos(kx)}{4\pi c}\hat{\w{x}},
\end{eqn}
where ${E_0=1.2\text{ statV/cm}}$ and ${k\lambda_D=0.8}$. To construct this perturbation, we first project the continuous 1-form field $\w{Y}_\text{cont}$ onto its Whitney form approximation, i.e. ${\w{Y}_\text{cont}\xrightarrow{\pi_h}\w{Y}=\w{y}\cdot\mM_1^{-1}\cdot\gv{\Lambda}^1}$. In particular, we solve for $\w{y}$ in Eq.~(\ref{AandYDefine}) such that the integrals of ${\w{Y}_\text{cont}}$ and $\w{Y}$ agree on edges of the discretized domain. This procedure yields a sinusoidal $\mu_\text{field}=\mG^T\w{y}$ as depicted in Fig.~\ref{MomentumMaps}.

From the particle side, electron momenta are initialized in phase space by randomly selecting their velocities $\dot{\w{X}}_\ell$ from a Maxwellian distribution of temperature ${T_e=5}$ keV. Taking ${\w{A}=0}$ at ${t=0}$, initial canonical momenta are therefore given by ${\w{P}_\ell=m_\ell\dot{\w{X}} _\ell}$. Electron positions are then initialized to be consistent with Gauss' law. More precisely, we demand that particle positions, in combination with the electric field perturbation, ensure the constancy of the total momentum map $\mu$. Following Eq.~(\ref{interpretMomMap}), a nonzero constant momentum map can be understood as a fixed, homogeneous, ion background charge, ${\mu\sim\rho_\text{ext}/c}$. Indeed, in a Landau damping simulation that takes only electrons to be dynamical, such a constant background charge constitutes the desired Gauss' law remainder.

We therefore optimize electron positions $\{\w{X}_\ell\}$ to ensure ${\mu_\text{particle}=\sum\limits_\ell\frac{\abs{e}}{c}\gv{\Lambda}^0(\w{X}_\ell)}$ closely satisfies
\begin{eqn}
\mu=\mu_\text{field}+\mu_\text{particle}\approx N_\text{ppc}\abs{e}/c,
\label{constantMu}
\end{eqn}
where ${\mu_\text{field}=\mG^T\w{y}}$ and ${N_\text{ppc}=40,000}$ denotes the number of particles per cell. In this way, the positive background charge, characterized by $\mu$, is homogeneous across the simulation domain and enforces quasineutrality with the dynamical electrons.

This optimization of ${\mu_\text{particle}}$ is carried out in two stages. First, electrons are randomly selected via rejection sampling \citep{von_neumann_various_1951} from the distribution
\begin{eqn}
n_e(x)=n_0\left[1+\frac{kE_0\sin(kx)}{4\pi\abs{e}n_0}\right],
\end{eqn}
which satisfies ${\nabla\cdot\w{E}=4\pi\abs{e}(n_0-n_e)}$. The electron positions ${\{\w{X}_\ell\}}$ are then further optimized using Nesterov accelerated gradient descent \citep{nesterov_method_1983} to minimize the objective function
\begin{eqn}
\argmin\limits_{\{\w{X}_\ell\}}\left|\mu-\frac{N_\text{ppc}\abs{e}}{c}\right|^2.
\end{eqn}
The resulting momentum map, plotted in red in Fig.~\ref{MomentumMaps}, thus defines a background charge that is homogeneous to a high degree of accuracy.

Note that an alternative initialization, undertaken for example by \citet{kraus_gempic:_2017}, reverses the above procedure by randomly generating electron positions first and then solving for $\w{y}$ to satisfy Gauss' law. This alternative is more straightforward computationally that the procedure described above, but it may afford less precision in the specification of the initial electric field perturbation, whenever such precision is desirable.

\begin{figure}
\begin{center}
\includegraphics[width=.8\linewidth]{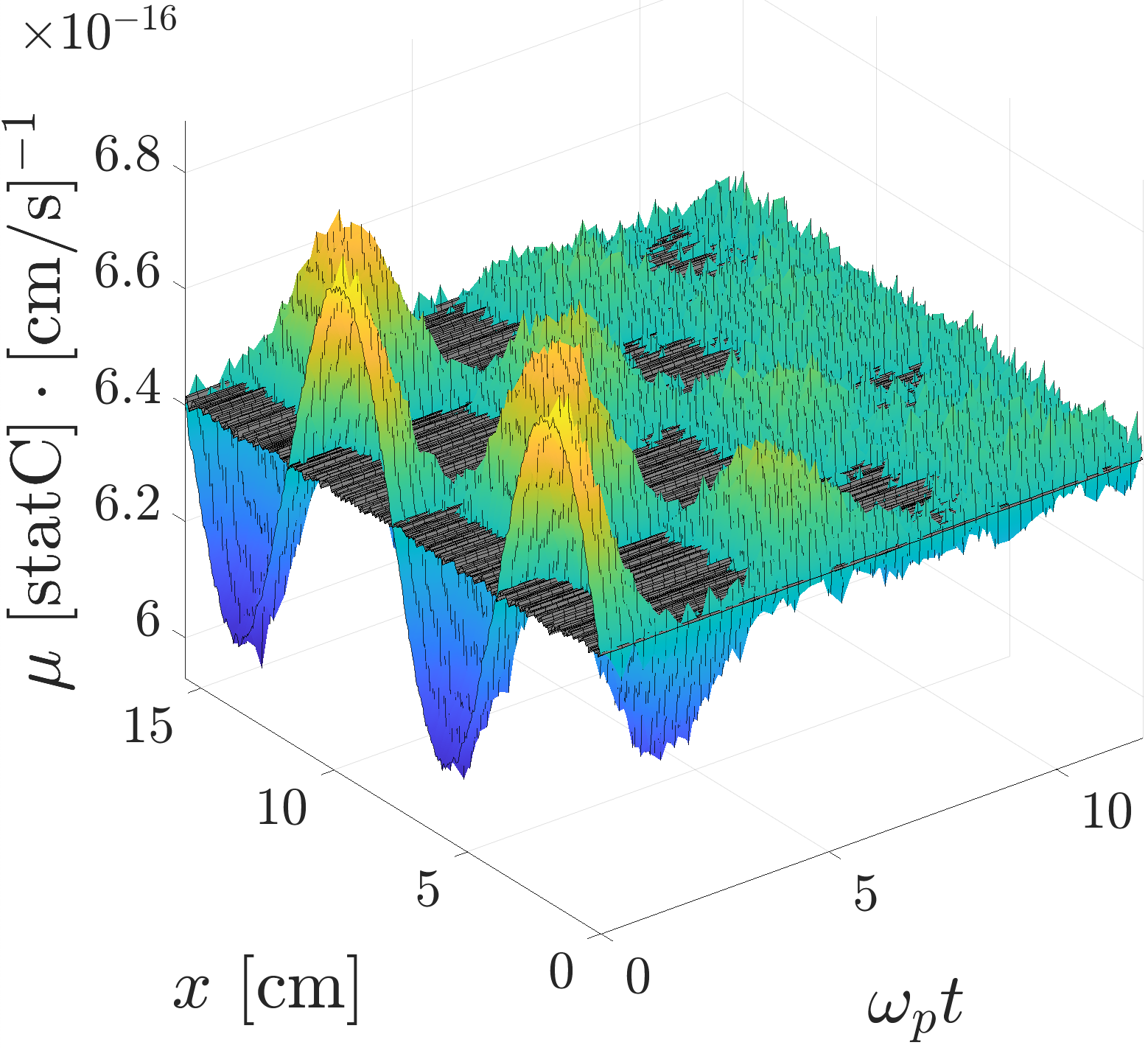}
\end{center}
\caption{With their initial conditions as depicted in Fig.~\ref{MomentumMaps}, the total momentum map $\mu$ (gray) is compared with $\mu_\text{particle}$ (multicolor) as the two functions evolve over time. Whereas $\mu_\text{particle}$ exhibits a decaying sinusoid consistent with Landau damping, $\mu$ remains constant to machine precision. The momentum map $\mu$ constitutes a phy sical representation of the fixed positive background charge implicit in the simulation.}
\label{muMap3D}
\end{figure}

With initialized fields and electrons, the simulation is then evolved using a first order Lie-Trotter splitting \citep{trotter_product_1959} derived from Eq.~(\ref{integratedEvolution}), in particular,
\begin{eqn}
\w{u}(t+\Delta)=\exp(\Delta H_\text{Kinetic}^x)\exp(\Delta H_\w{Y})\exp(\Delta H_\w{A})\w{u}(t)
\end{eqn}
where ${\w{u}(t)}$ denotes the simulation state at time $t$---i.e. ${\w{u}=m_d\in M_d}$, a point in phase space as defined in Eq.~(\ref{phaseSpacePt}).

In Fig.~\ref{muMap3D}, we first examine the evolution of the momentum map throughout the simulation domain. We note that, while $\mu_\text{particle}$ (multicolor) exhibits an oscillation and decay consistent with Landau damping, $\mu$ (gray) remains constant over time, consistent with the conservation of the momentum map.

\begin{figure}
\begin{center}
\includegraphics[width=\linewidth]{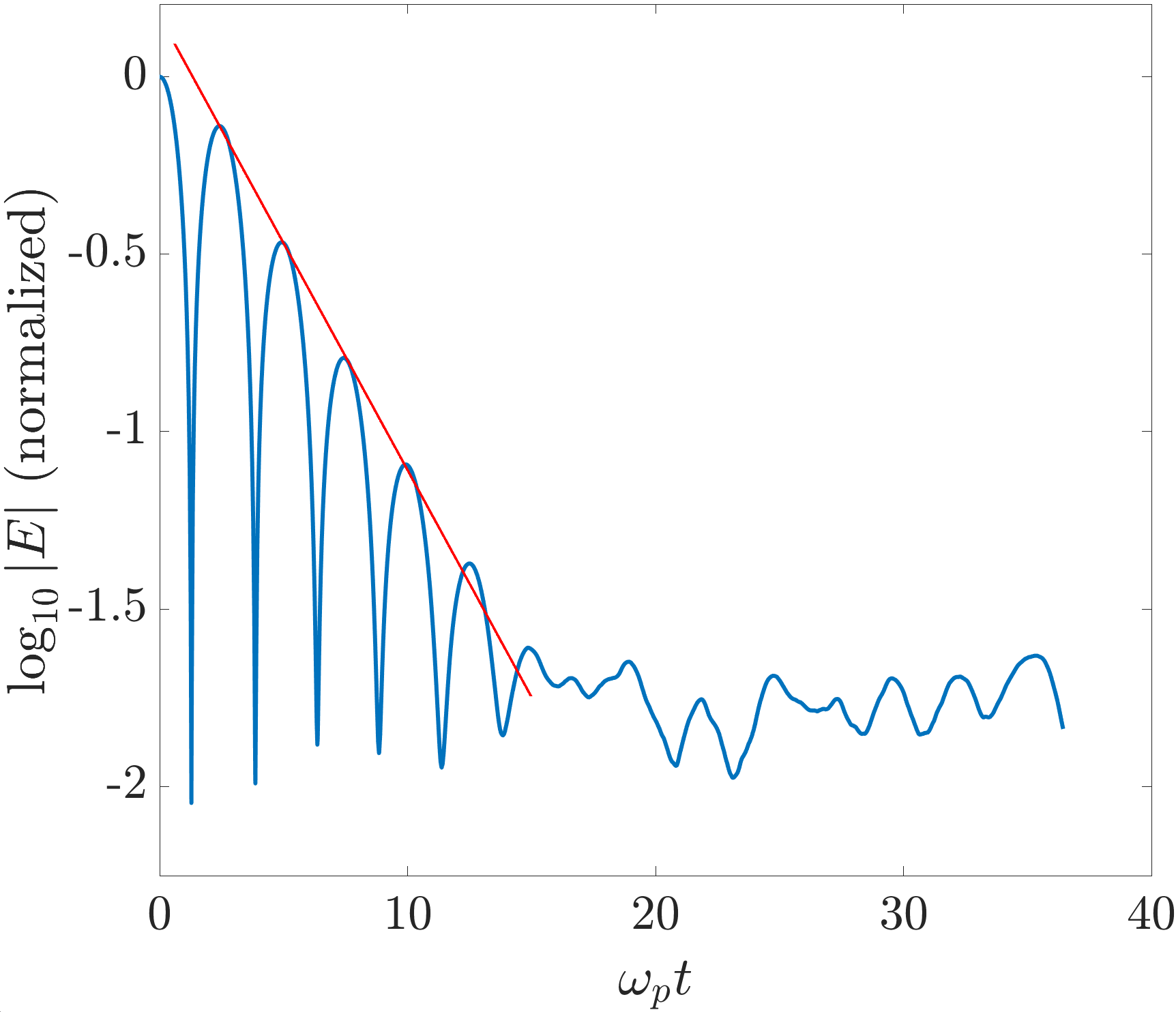}
\end{center}
\caption{The evolution of an electrostatic wave over time is simulated with a first order Lie-Trotter splitting \citep{trotter_product_1959} of Eq.~(\ref{integratedEvolution}). The blue time series denotes the (normalized) log modulus of the electric field ${\w{E}=-4\pi c\w{Y}}$, where ${\abs{\w{E}}}$ is computed over the simulation domain by the ${L^2\Lambda^1}$ norm. The theoretical Landau damping rate of the wave in a Maxwellian plasma is depicted as a red line, decaying at a rate of ${\omega_i=\frac{\omega_p}{\kappa^3}\sqrt{\frac{\pi}{8}}\exp\left(-\frac{1+3\kappa^2}{2\kappa^2}\right)}$ for ${\kappa=k\lambda_D}$.}
\label{LandauDampPlot}
\end{figure}

\begin{figure}
\begin{center}
\includegraphics[width=\linewidth]{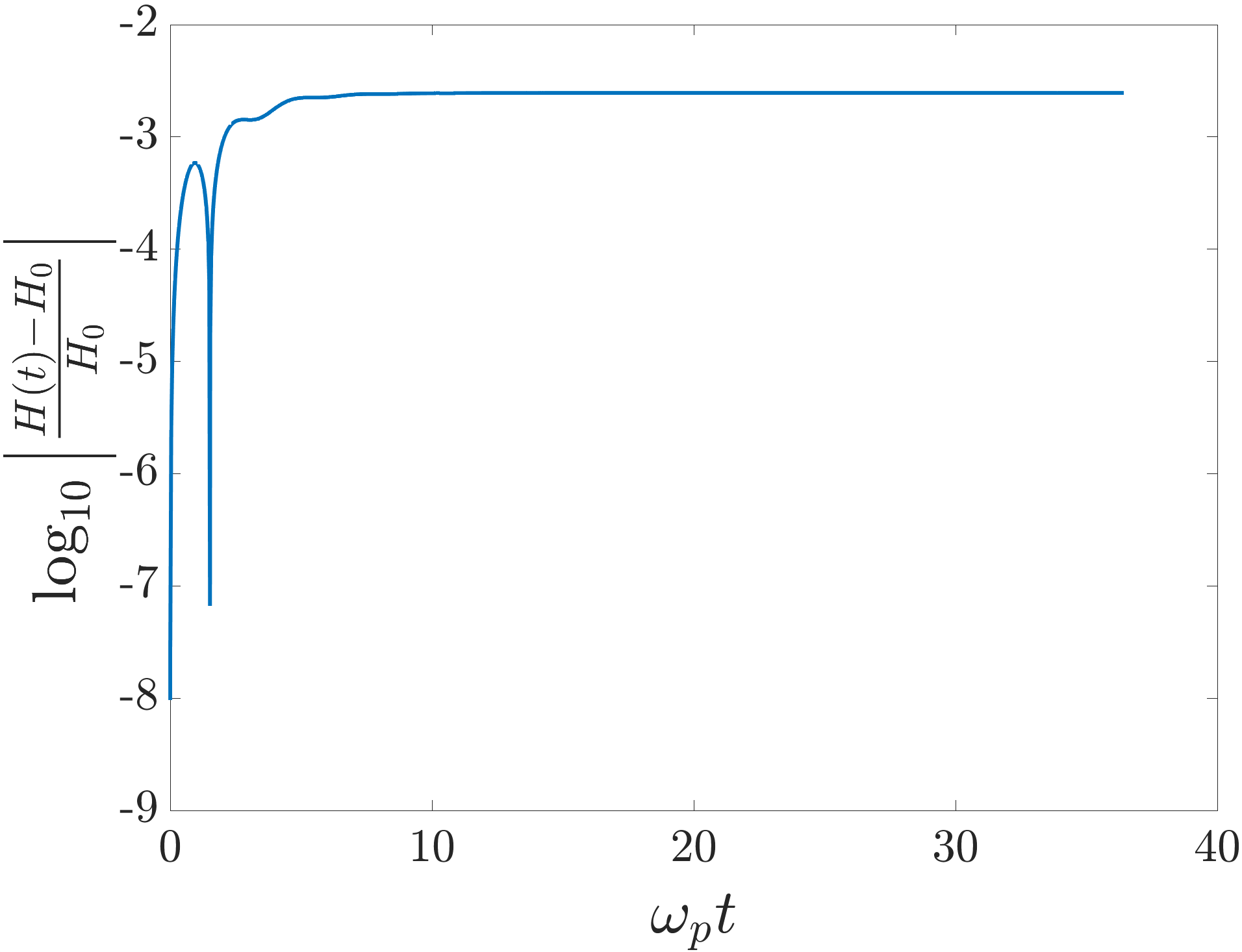}
\end{center}
\caption{The log error in the total energy of a first order Lie-Trotter splitting \citep{trotter_product_1959} Landau damping simulation.}
\label{EnergyError}
\end{figure}

To compare this simulation with theory, the evolution of the (normalized) electric field is plotted in Fig.~\ref{LandauDampPlot}, (which may be compared with Fig. 2 of \citet{xiao_explicit_2015}). The results agree with a theoretical expectation of (i) electrostatic Langmuir wave oscillation at a frequency ${\omega_p=3.0\times10^9\text{ rad/s}}$, and (ii) Landau damping at a decay rate ${\omega_i=\frac{\omega_p}{\kappa^3}\sqrt{\frac{\pi}{8}}\exp\left(-\frac{1+3\kappa^2}{2\kappa^2}\right)=3.9\times10^8\text{ rad/s}}$, where ${\kappa=k\lambda_D}$. Furthermore, as is characteristic of a symplectic algorithm, the error in the total energy, measured by $H_\text{VM}$ of Eq.~(\ref{discHamiltonian}), is well bounded throughout the simulation. This error is plotted in Fig.~\ref{EnergyError}.

Having demonstrated the canonical finite element formalism's efficacy in simulating an electrostatic problem, we next consider the electromagnetic simulation of the Weibel instability. Before that, however, we introduce the ``canonical 1X2P phase space" in which we will conduct our simulation.

\subsection{Canonical 1X2P Phase Space\label{1X2PSect}}

Let us consider the restriction of phase space to one spatial dimension and two dimensions of canonical momentum. Despite its computational efficiency, the 1X2P setting is capable of simulating a number of nontrivial electromagnetic problems. Our development here of 1X2P phase space essentially adapts for canonical coordinates the techniques of \citet{crouseilles2015Hamiltonian} and \citet{kraus_gempic:_2017}.

To characterize 1X2P, we must reflect the lack of $y$ dependence in the finite element expansions of our fields. In this setting, we therefore define $\w{A}$ and $\w{Y}$ by analogy with Eq.~(\ref{AandYDefine}) as
\begin{eqn}
\w{A}&=\w{a}_x\cdot\gv{\Lambda}^1(x)+\w{a}_y\cdot\gv{\Lambda}^0(x)\\
\w{Y}&=\w{y}_x\cdot\mM_1^{-1}\cdot\gv{\Lambda}^1(x)+\w{y}_y\cdot\mM_0^{-1}\cdot\gv{\Lambda}^0(x).
\label{ExpansionAY1X2P}
\end{eqn}
Here, ${\w{a}_x}$ denotes a vector of coefficients that pair only with the finite element 1-form basis ${\gv{\Lambda}^1(x)}$---a basis in 1X2P which has only $x$-components and whose spatial dependence is only one-dimensional. We further note that $\w{a}_y$ is defined as the coefficients of a \emph{0-form}. Since there is no way to associate 1-forms with $\hat{y}$-directed edges in the 1X2P setting, 0-forms are the most natural representation of the $y$ components of $\w{A}$. This becomes especially clear when examining finite elements restricted to have spatial $x$-dependence. The expansion of $\w{Y}$ in Eq.~(\ref{ExpansionAY1X2P}) follows similarly, with appropriate mass matrices computed  by integrals over the one-dimensional domain.

Turning now to the characterization of particle phase space, we retain two components of the canonical momentum, $P_{\ell x}$ and $P_{\ell y}$. Only one component of spatial dependence is retained, which we shall denote $X_\ell$.

The appropriate Hamiltonian is then computed by restricting Eq.~(\ref{discHamiltonian}) as follows:
\begin{eqn}
H_\text{1X2P}=\frac{1}{8\pi}&\bigg((4\pi c)^2\Big[\w{y}_x\cdot\mM_1^{-1}\cdot\w{y}_x+\w{y}_y\cdot\mM_0^{-1}\cdot\w{y}_y\Big]+\w{a}_y\cdot\mG^T\mM_1\mG\cdot\w{a}_y\bigg)\\
&+\sum\limits_{\ell=1}^L\frac{1}{2m_\ell}\left(\Big|P_{\ell x}-\frac{q_\ell}{c}\w{a}_x\cdot\gv{\Lambda}^1(X_\ell)\Big|^2+\Big|P_{\ell y}-\frac{q_\ell}{c}\w{a}_y\cdot\gv{\Lambda}^0(X_\ell)\Big|^2\right).
\label{discHamiltonian1X2P}
\end{eqn}
We note that the term ${\w{a}_x\cdot\mC^T\mM_2\mC\cdot\w{a}_x}$ is absent; while it would ordinarily arise from the magnetic energy $|\md\w{A}|^2$, $\md$ annihilates 1-forms in the 1X2P setting. Indeed, the magnetic field in the 1X2P formalism is given simply by ${\w{B}=\md\w{A}=\w{a}_y\cdot\md\gv{\Lambda}^0=\mG\w{a}_y\cdot\gv{\Lambda}^1}$.

The Poisson bracket of the restricted phase space follows by simply omitting the inapplicable terms of Eq.~(\ref{discretePoissonBracket}). In particular:
\begin{eqn}
\{F,G\}_\text{1X2P}&=\left(\pd{F}{\w{a}_x}\cdot\pd{G}{\w{y}_x}+\pd{F}{\w{a}_y}\cdot\pd{G}{\w{y}_y}+\sum\limits_{\ell=1}^L\frac{\partial F}{\partial X_\ell}\frac{\partial G}{\partial P_{\ell x}}\right)-(F\leftrightarrow G).
\label{discretePoissonBracket1X2P}
\end{eqn}

We observe that, despite its simplicity, the 1X2P phase space retains the gauge structure of the original problem. In particular we note its gauge symmetry 
\begin{eqn}
\Phi_{\exp(\w{s})}\left(\begin{matrix}
\w{a}_x\\
P_{\ell x}
\end{matrix}\right)=\left(\begin{matrix}
\w{a}_x+\mG\w{s}_x\\
P_{\ell x}+\frac{q_\ell}{c}\mG\w{s}_x\cdot\gv{\Lambda}^1(X_\ell)_x\end{matrix}\right)
\label{gaugeTransf1X2P}
\end{eqn}
and the resulting momentum map
\begin{eqn}
\mu_\text{1X2P}=\mG^T\w{y}_x-\sum\limits_{\ell=1}^L\frac{q_\ell}{c}\gv{\Lambda}^0(X_\ell).
\end{eqn}

Finally, we derive the equations of motion in the 1X2P setting, which are straightforward reexpressions of their full phase space counterparts in Eq.~(\ref{theEOM}):
\begin{eqn}
\dot{X}_\ell&=\{X_\ell,H_\text{1X2P}\}=\frac{1}{m_\ell}\Big(P_{\ell x}-\frac{q_\ell}{c}\w{a}_x\cdot\gv{\Lambda}^1(X_\ell)\Big)\\
\dot{P}_{\ell x}&=\{P_{\ell x},H_\text{1X2P}\}=\frac{q_\ell}{c}\Big[\dot{X}_\ell\w{a}_x\cdot\left(\partial_{X_\ell}\gv{\Lambda}^1(X_\ell)\right)+\dot{Y}_\ell\w{a}_y\cdot\left(\partial_{X_\ell}\gv{\Lambda}^0(X_\ell)\right)\Big]\\
\dot{\w{a}}_x&=\{\w{a}_x,H_\text{1X2P}\}=4\pi c^2\mM_1^{-1}\cdot\w{y}_x\\
\dot{\w{a}}_y&=\{\w{a}_y,H_\text{1X2P}\}=4\pi c^2\mM_0^{-1}\cdot\w{y}_y\\
\dot{\w{y}}_x&=\{\w{y}_x,H_\text{1X2P}\}=\sum\limits_{\ell=1}^L\frac{q_\ell}{c}\dot{X}_\ell\gv{\Lambda}^1(X_\ell)\\
\dot{\w{y}}_y&=\{\w{y}_y,H_\text{1X2P}\}=\sum\limits_{\ell=1}^L\frac{q_\ell}{c}\dot{Y}_\ell\gv{\Lambda}^0(X_\ell)-\frac{1}{4\pi}\mG^T\mM_1\mG\cdot\w{a}_y.
\label{continuousTime1X2PEOM}
\end{eqn}
Here, with an abuse of notation that ignores the non-existence of $Y_\ell$, we define
\begin{eqn}\dot{Y}_\ell=\frac{1}{m_\ell}\left(P_{\ell y}-\w{a}_y\cdot\gv{\Lambda}^0(X_\ell)\right).
\end{eqn}
We observe that $P_{\ell y}$ is conserved in Eq.~(\ref{continuousTime1X2PEOM}), which can be viewed as a consequence of the independence from $Y_\ell$ of the 1X2P formulation.

We further note that a gauge-compatible splitting of $H_\text{1X2P}$ is readily defined by the following four sub-Hamiltonians, in agreement with Eq.~(\ref{splitHamiltonian}):
\begin{eqn}
H_\text{1X2P}=H_\w{A}+H_\w{Y}&+H^x_\text{Kinetic}+H^y_\text{Kinetic}\\
\text{where}~~~~~~~~H_\w{A}&=\frac{1}{8\pi}\w{a}_y\cdot\mG^T\mM_1\mG\cdot\w{a}_y\\
H_\w{Y}&=\frac{1}{8\pi}\bigg((4\pi c)^2\Big[\w{y}_x\cdot\mM_1^{-1}\cdot\w{y}_x+\w{y}_y\cdot\mM_0^{-1}\cdot\w{y}_y\Big]\bigg)\\
H^x_\text{Kinetic}&=\sum\limits_{\ell=1}^L\frac{1}{2m_\ell}\Big|P_{\ell x}-\frac{q_\ell}{c}\w{a}_x\cdot\gv{\Lambda}^1(X_\ell)\Big|^2\\
H^y_\text{Kinetic}&=\sum\limits_{\ell=1}^L\frac{1}{2m_\ell}\Big|P_{\ell y}-\frac{q_\ell}{c}\w{a}_y\cdot\gv{\Lambda}^0(X_\ell)\Big|^2.
\label{discHamiltonian1X2PSplit}
\end{eqn}
Using the Poisson bracket ${\{\cdot,\cdot\}_\text{1X2P}}$ of Eq.~(\ref{discretePoissonBracket1X2P}), we thereby derive the following subsystem equations of motion:
\begin{eqn}
H_\w{A}:
\begin{cases}
\dot{\w{y}}_y=-\frac{1}{4\pi}\mG^T\mM_1\mG\cdot\w{a}_y
\end{cases}
%%%%%
\hspace{105pt}
%%%%%
H_\w{Y}:
\begin{cases}
\dot{\w{a}}_x=4\pi c^2\mM_1^{-1}\cdot\w{y}_x\vspace{4pt}\\
\dot{\w{a}}_y=4\pi c^2\mM_0^{-1}\cdot\w{y}_y\vspace{4pt}
\end{cases}\hspace{40pt}\\
%%%%%
\hspace{-30pt}H_{\text{Kinetic}}^x:
\begin{cases}
\dot{X}_\ell&=\frac{1}{m_\ell}\Big(P_{\ell x}-\frac{q_\ell}{c}\w{a}_x\cdot\gv{\Lambda}^1(\w{X}_\ell)\Big)\vspace{4pt}\\
\dot{P}_{\ell x}&=\frac{q_\ell}{c}\dot{X}_\ell\w{a}_x\cdot\left(\partial_{X_\ell}\gv{\Lambda}^1(X_\ell)\right)\vspace{4pt}\\
\dot{\w{y}}_x&=\sum\limits_{\ell=1}^L\frac{q_\ell}{c}\dot{X}_\ell\gv{\Lambda}^1(X_\ell)
\end{cases}
%%%%%
\hspace{40pt}
%%%%%
H_{\text{Kinetic}}^y:
\begin{cases}
\dot{P}_{\ell x}&=\frac{q_\ell}{c}\dot{Y}_\ell\w{a}_y\cdot\left(\partial_{X_\ell}\gv{\Lambda}^0(X_\ell)\right)\vspace{4pt}\\
\dot{\w{y}}_y&=\sum\limits_{\ell=1}^L\frac{q_\ell}{c}\dot{Y}_\ell\gv{\Lambda}^0(X_\ell).
\end{cases}
\label{SubsystemEOM1X2P}
\end{eqn}
Since these equations of motion are---as  Eq.~(\ref{integratedEvolution})---exactly solvable, and since the sub-Hamiltonians of Eq.~(\ref{discHamiltonian1X2PSplit}) preserve the gauge symmetry of Eq.~(\ref{gaugeTransf1X2P}), it is clear that Eq.~(\ref{SubsystemEOM1X2P}) defines a gauge-compatible splitting method that exactly preserves ${\mu_\text{1X2P}}$.

\subsection{Weibel Instability\label{WeibelInstability}}

We now apply the 1X2P formulation defined above to the simulation of the Weibel instability. We initialize the simulation in close agreement with the parametrization of the Weibel instability simulations of \citet{crouseilles2015Hamiltonian} and \citet{kraus_gempic:_2017}. Specifically, in a periodic domain of 64 cells we simulate ${N_e=100,032}$ electrons. Continuing to work in Gaussian units, we take perturbation wavenumber ${k=1.25\frac{\omega_p}{c}}$ and simulation domain length ${L=2\pi/k}$. We seed a magnetic perturbation by defining the vector potential to be 
\begin{eqn}
\w{A}&=\frac{B_0}{k}\sin(kx)\hat{y}
\end{eqn}
where ${B_0=-5.7\text{mG}}$. We sample initial electron velocities from the anisotropic distribution
\begin{eqn}
f_e(x,v_x,v_y)=\frac{1}{2\pi\sigma_x\sigma_y}\exp\left(-\left[\frac{v_x^2}{2\sigma_x^2}+\frac{v_y^2}{2\sigma_y^2}\right]\right)
\end{eqn}
where ${\sigma_x=0.02c/\sqrt{2}}$ and ${\sigma_y=\sqrt{12}\sigma_x}$. 
To initialize the initial electron distribution to be independent of $x$ (i.e., spatially uniform) as above, we evenly space electrons throughout the simulation domain at  separation of $L/N_e$. $\w{Y}$ is correspondingly initialized to zero. $\w{A}$ and $\w{Y}$ are again modeled using Whitney (0- and 1-) forms.

\begin{figure}
\begin{center}
\includegraphics[width=\linewidth]{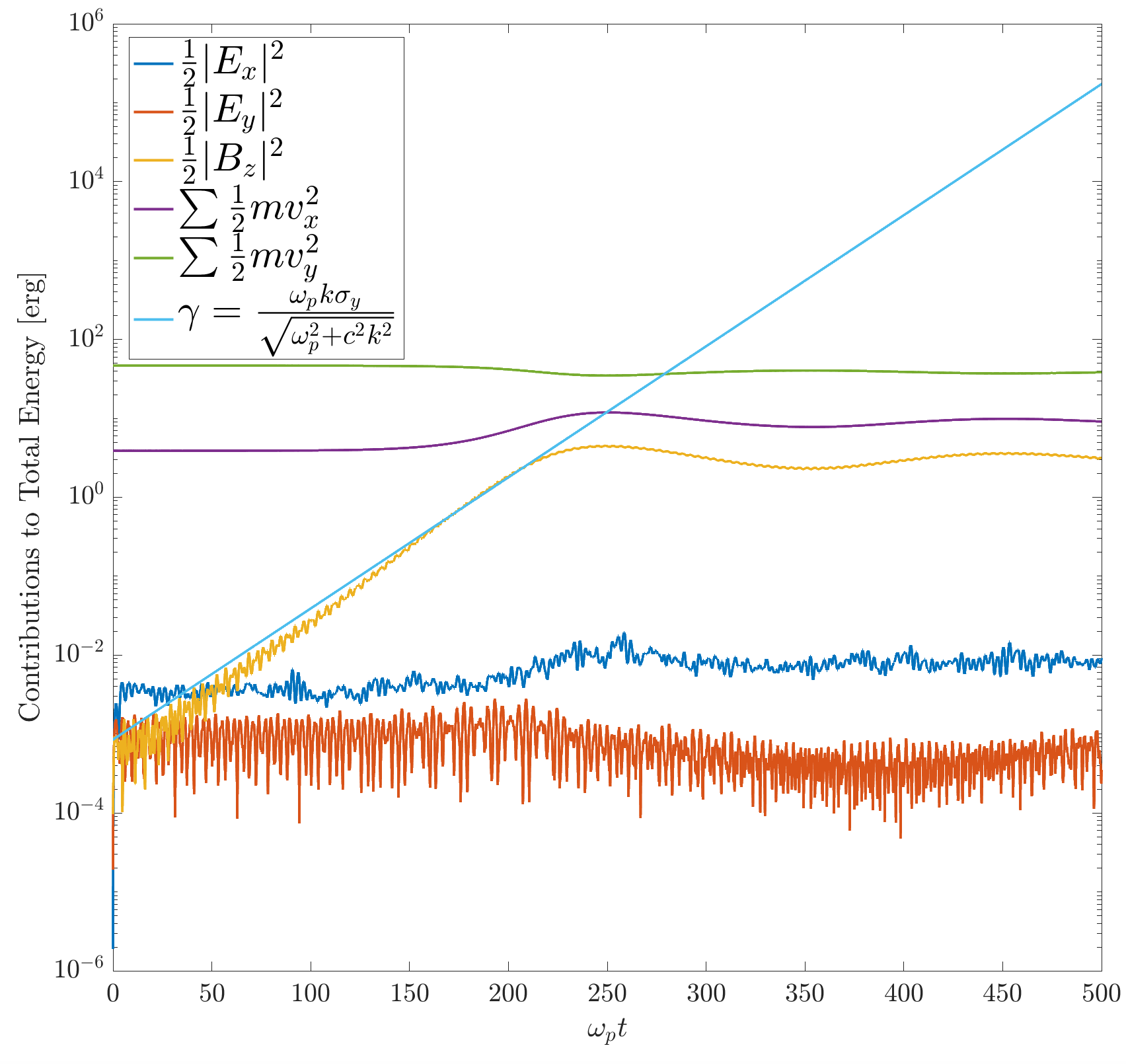}
\end{center}
\caption{The growth rate of magnetic field energy closely approximates the analytic model of the Weibel instability. As the magnetic field ramps up, the anisotropy of electron velocity is reduced. This plot may be compared with Fig.~1 of \citet{kraus_gempic:_2017}.}
\label{WeibelEvolution}
\end{figure}

\begin{figure}
\begin{center}
\includegraphics[width=\linewidth]{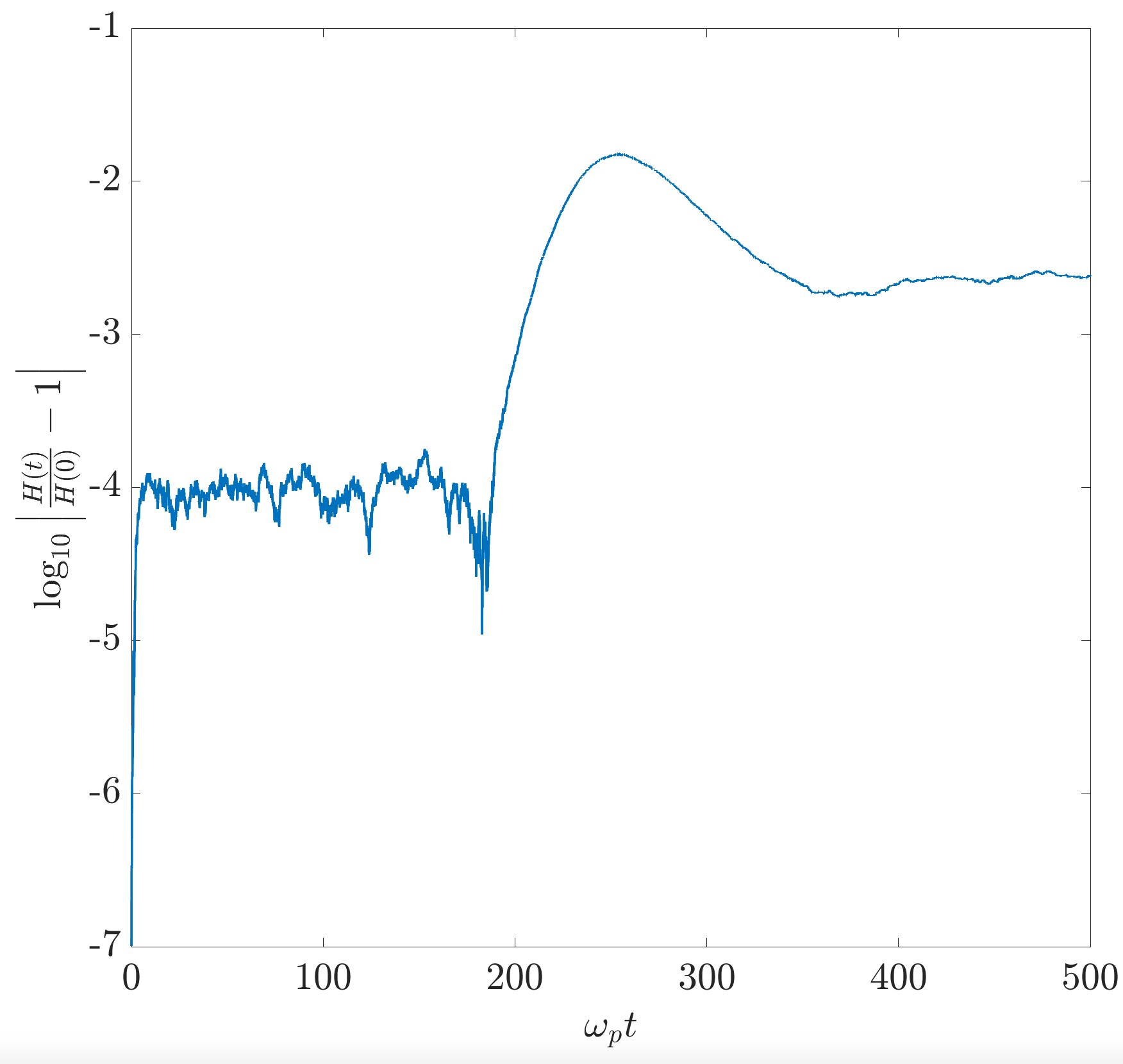}
\end{center}
\caption{The log error in the total energy of a first order Lie-Trotter splitting \citep{trotter_product_1959} Weibel instability simulation \citep{weibel_spontaneously_1959}.}
\label{WeibelEnergyError}
\end{figure}

Finally, simulating the Weibel instability with a Lie-Trotter splitting of stepsize $1/40\omega_p$ yields the evolution plotted in Fig.~\ref{WeibelEvolution}. The magnetic field growth rate is in strong agreement with the analytic prediction \citep{weibel_spontaneously_1959},
\begin{eqn}
\gamma=\frac{\omega_pk\sigma_y}{\sqrt{\omega_p^2+c^2k^2}}.
\end{eqn}
Lastly, the total energy is also well bounded over the lifetime of the Weibel instability simulation, as depicted in Fig.~\ref{WeibelEnergyError}.

\section{Conclusion\label{ConcludeSection}}

We have derived a canonical Poisson structure in Eq.~(\ref{discretePoissonBracket}) for the Vlasov-Maxwell system and constructed its Hamiltonian in Eq.~(\ref{discHamiltonian}) in the formalism of finite element exterior calculus. Its gauge symmetry was studied to systematically derive the corresponding charge-conserving momentum map of Eq.~(\ref{MomMapGaussLaw}). The resulting PIC algorithm of Eq.~(\ref{integratedEvolution}) was demonstrated to be a gauge-compatible splitting method that preserves the momentum map to machine precision over the full simulation domain.

We have seen in Eq.~(\ref{interpretMomMap}) how the momentum map may be regarded as an external fixed charge in a PIC simulation. Using this interpretation, we optimized initial conditions for a Landau damping simulation that modeled a homogeneous positive fixed background, as depicted in Fig.~\ref{MomentumMaps}. We explicitly demonstrated the preservation of this momentum map, as seen in Fig.~\ref{muMap3D}.

The initialization procedure we described may be useful in future structure-preserving PIC simulations that require the precise initial specification of electric fields and background charge. Such a technique might be advantageous, for example, in the simulation of plasma interactions with charged plasma-facing components.

We further explored the efficacy of our gauge-compatible splitting algorithm by examining its restriction to 1X2P phase space, in which setting we simulated the Weibel instability. We demonstrated that the gauge structure of the full phase space has a counterpart in this sparser setting, and we accurately modeled the analytic growth rate of the Weibel instability in our simulation.

The flexibility of finite element exterior calculus makes the algorithms defined in this paper significantly generalizable as well. For example, the PIC algorithm of this paper may be adapted to simulations on an unstructured mesh, including perhaps those defined in curvilinear coordinates.

However, it is worth noting an important drawback to the use of higher order finite element spaces in the foregoing algorithms, namely, that their inverse mass matrices (e.g. $\mM_1^{-1}$ in Eq.~\ref{SubsystemEOM}) are not, in general, sparse. Dense inverse mass matrices necessitate global communication in every timestep of a simulation, significantly reducing the benefit of parallelization. Overcoming such a limitation would facilitate scalable, higher order, and higher dimensional simulations---an effort that will be worthy of future study.

\section*{Acknowledgements}
The authors thank the referees for their advice in evaluating this article.

\section*{Funding}
This research was supported by the U.S. Department of Energy (DE-AC02-09CH11466). This research was further supported by the U.S. Department of Energy Fusion Energy Sciences Postdoctoral Research Program administered by the Oak Ridge Institute for Science and Education (ORISE) for the DOE. ORISE is managed by Oak Ridge Associated Universities (ORAU) under DOE contract number DE-SC0014664. All opinions expressed in this paper are the authors' and do not necessarily reflect the policies and views of DOE, ORAU, or ORISE.

\section*{Declaration of Interests}
The authors report no conflict of interest.

\section*{Author ORCID}
A.S. Glasser, \url{https://orcid.org/0000-0001-8357-8278}; H. Qin \url{https://orcid.org/
0000-0003-0304-3762}

\section*{Author Contributions} The authors contributed equally to theoretical development. A.S.G. performed the derivations and simulations.

% Create the reference section using BibTeX: %
%\bibliography{allrefs.bib}
\bibliographystyle{jpp}
\bibliography{allrefs}

\end{document}